\documentclass[%
reprint,
superscriptaddress,
amsmath,amssymb,
aps,
floatfix]{revtex4-2}

\usepackage{graphicx}
\usepackage{dcolumn}
\usepackage{bm}
\usepackage[hidelinks,colorlinks=true,linkcolor=blue,citecolor=blue,urlcolor=blue]{hyperref}

\begin{document}

\preprint{Kraichnan_model}

\title{Phase-space entropy cascade and irreversibility of stochastic heating in nearly collisionless plasma turbulence}

\author{Michael L. Nastac}
\email{michael.nastac@physics.ox.ac.uk}
\affiliation{Rudolf Peierls Centre for Theoretical Physics, University of Oxford, Clarendon Laboratory, Parks Road, Oxford OX1 3PU, UK}
\affiliation{St. John’s College, Oxford OX1 3JP, UK}
\affiliation{Institute for Research in Electronics and Applied Physics, University of Maryland, College Park, MD, 20742, USA}

\author{Robert J. Ewart}
\affiliation{Rudolf Peierls Centre for Theoretical Physics, University of Oxford, Clarendon Laboratory, Parks Road, Oxford OX1 3PU, UK}
\affiliation{Balliol College, Oxford, OX1 3BJ, UK}

\author{Wrick Sengupta}
\affiliation{Department of Astrophysical Sciences, Princeton University, Princeton, New Jersey 08543, USA}
\affiliation{Princeton Plasma Physics Laboratory, Princeton, New Jersey 08540, USA}

\author{Alexander~A.~Schekochihin}
\affiliation{Rudolf Peierls Centre for Theoretical Physics, University of Oxford, Clarendon Laboratory, Parks Road, Oxford OX1 3PU, UK}
\affiliation{Merton College, Oxford OX1 4JD, UK}

\author{Michael Barnes}
\affiliation{Rudolf Peierls Centre for Theoretical Physics, University of Oxford, Clarendon Laboratory, Parks Road, Oxford OX1 3PU, UK}
\affiliation{University College, Oxford OX1 4BH, UK}

\author{William D. Dorland}
\affiliation{Rudolf Peierls Centre for Theoretical Physics, University of Oxford, Clarendon Laboratory, Parks Road, Oxford OX1 3PU, UK}
\affiliation{Institute for Research in Electronics and Applied Physics, University of Maryland, College Park, MD, 20742, USA}
\affiliation{Department of Physics, University of Maryland, College Park, Maryland 20740, USA}

\date{\today}

\begin{abstract}

We consider a nearly collisionless plasma consisting of a species of `test particles' in one spatial and one velocity dimension, stirred by an externally imposed stochastic electric field---a kinetic analogue of the Kraichnan model of passive advection. The mean effect on the particle distribution function is turbulent diffusion in velocity space—known as stochastic heating. Accompanying this heating is the generation of fine-scale structure in the distribution function, which we characterize with the collisionless (Casimir) invariant $C_2 \propto \iint dx dv \, \langle f^2 \rangle$—a quantity that here plays the role of (negative) entropy of the distribution function. We find that $C_2$ is transferred from large scales to small scales in both position and velocity space via a phase-space cascade enabled by both particle streaming and nonlinear interactions between particles and the stochastic electric field. We compute the steady-state fluxes and spectrum of $C_2$ in Fourier space, with $k$ and $s$ denoting spatial and velocity wavenumbers, respectively. In our model, the nonlinearity in the evolution equation for the spectrum turns into a fractional Laplacian operator in $k$ space, leading to anomalous diffusion. Whereas even the linear phase mixing alone would lead to a constant flux of $C_2$ to high $s$ (towards the collisional dissipation range) at every $k$, the nonlinearity accelerates this cascade by intertwining velocity and position space so that the flux of $C_2$ is to both high $k$ and high $s$ simultaneously. Integrating over velocity (spatial) wavenumbers, the $k$-space ($s$-space) flux of $C_2$ is constant down to a dissipation length (velocity) scale that tends to zero as the collision frequency does, even though the rate of collisional dissipation remains finite. The resulting spectrum in the inertial range is a self-similar function in the $(k,s)$ plane, with power-law asymptotics at large $k$ and $s$. Our model is fully analytically solvable, but the asymptotic scalings of the spectrum can also be found via a simple phenomenological theory whose key assumption is that the cascade is governed by a `critical balance' in phase space between the linear and nonlinear time scales. We argue that stochastic heating is made irreversible by this entropy cascade and that, while collisional dissipation accessed via phase mixing occurs only at small spatial scales rather than at every scale as it would in a linear system, the cascade makes phase mixing even more effective overall in the nonlinear regime than in the linear one.

\end{abstract}

\maketitle

\section{Introduction} \label{intro}
Understanding the nature of turbulent cascades in nearly collisionless space and astrophysical plasmas is an outstanding problem \cite{marsch2006kinetic,schekochihin2009astrophysical,zimbardo2010magnetic,chen2016recent,howes2017prospectus,verscharen2019multi,prop_solar_corona_wind} with a diverse range of applications, from solving the coronal-heating problem \cite{prop_solar_corona_wind} to interpreting radiation emission from accretion disks around black holes \cite{quataert1999turbulence,chael2018role, akiyama2019first}. A major difficulty distinguishing such turbulence from its fluid counterparts lies in the fact that fluctuations evolve in the six-dimensional phase space of single-particle positions and velocities. Dissipation (in the sense of irreversible entropy production) occurs via particle collisions \cite{Boltzmann,landau1936transport}, which in a nearly collisionless plasma are activated only when the particle distribution function develops large gradients in velocity space.

Turbulent dissipation in nearly collisionless plasmas is often considered from an energetics perspective. This point of view focuses on the cascade of bulk kinetic and electromagnetic energy from large to small spatial scales (see, e.g., \cite{parashar2015turbulent,chen2016recent,matthaeus2020pathways,roy2022turbulent,marino2023scaling} and references therein) and the physical processes, such as magnetic reconnection \cite{birn2001geospace,yamada2010magnetic,lazarian20203d,ji2022magnetic} and wave-particle interactions \cite{marsch2006kinetic,verscharen2019multi}, that convert this energy into the internal energy of the plasma. Whilst energy-cascade and transfer mechanisms are important, they provide a fundamentally incomplete picture of turbulent dissipation. Without collisions, entropy is formally conserved (along with an infinity of Casimir invariants \cite{ye1992action}), and any transfer of energy between particles and fields is formally reversible.

It was realized by \cite{krommes1994role,krommes1999thermostatted} that entropy production in the long-time limit of a nearly collisionless plasma must remain finite even as the collision frequency $\nu$ tends to zero. This idea crystallized in \cite{schekochihin2008gyrokinetic,schekochihin2009astrophysical,tatsuno2009nonlinear,plunk2010two}, where the notion of entropy cascade in the context of gyrokinetics was introduced. Below the Larmor scale, in a phase-space inertial range between the injection range at large scales and the collisional dissipation range at small scales, the (negative) entropy of the perturbed distribution function cascades in both position and velocity space via a nonlinear perpendicular phase-mixing mechanism \cite{dorland1993gyrofluid}. Because this is a constant-flux cascade, the turbulent heating rate in a gyrokinetic plasma is finite and independent of $\nu$ even as $\nu \rightarrow 0^{+}$, analogous to so-called dissipative anomalies \cite{frisch1995turbulence,eyink2006onsager} in hydrodynamics, where viscous dissipation is finite in infinite-Reynolds-number turbulence.

Entropy cascade outside the gyrokinetic approximation is a frontier topic just beginning to be explored \cite{servidio2017magnetospheric,cerri2018dual,pezzi2018velocity,eyink2018cascades,celebre2023}. In this paper, we study turbulent dissipation and entropy cascade via an analytically solvable model introduced by \cite{adkins2018solvable}. We consider the `1D-1V' electrostatic, ~{full-$f$ Vlasov} equation with a model diffusive collision operator for a test-particle species, where instead of the electric field being self-consistently determined by Poisson's equation, it is externally determined to be a stochastic Gaussian, white-noise source with a specified spatial correlation function. Physically, this model represents the evolution of a low-density minority species in a multi-component plasma whose dielectric response is dominated by the other, more abundant species. This model is the plasma-kinetic analogue of the Kraichnan \cite{kraichnan1968small} model of passive advection, where a scalar field \footnote{Or a vector field, such as a magnetic field, in which case the model is named after Kazantsev, who, in the year before Kraichnan, pursued the same type of approximation for the kinematic-dynamo problem \cite{kazantsev1968enhancement, rincon_2019}.}, such as temperature or concentration of dye, is passively advected by an externally determined random flow, allowing for analytical calculations of the passive-scalar statistics. The Kraichnan model has been dubbed the `Ising model' \cite{eyink2011robert} of fluid turbulence because it is a solvable model that exhibits many properties also present in real systems, and so serves as a theoretical laboratory to study turbulence \cite{falkovich2001particles}. It is in this spirit that we investigate our solvable model of kinetic plasma turbulence.

We decompose the distribution function into its mean and fluctuating parts, $f =\langle f \rangle + \delta \! f$, where $\langle ... \rangle$ denotes ensemble averaging over realizations of the random electric field. In our model, particles are stochastically accelerated by the electric field and undergo random walks in velocity space, resulting in bulk heating of $\langle f \rangle$ \cite{sturrock1966stochastic}. In the context of magnetized plasmas, this phenomenon is often referred to as stochastic heating \cite{chandran2010perpendicular,verscharen2019multi,cerri2021stochastic}. Accompanying this heating is the generation of velocity and spatial structure in the perturbed distribution function $\delta \! f$ via linear phase mixing and nonlinear interactions between particles and the turbulent electric field \cite{adkins2018solvable}.

We characterize this structure via the collisionless (Casimir) invariant $C_2 = (1/L) \iint dx dv \, \langle f^2 \rangle/2 $, where $L$ is the system size. $C_2$ has been considered before as a measure of phase-space structure and as a cascaded quantity in kinetic plasma turbulence \cite{diamond2010modern,kosuga2011relaxation, lesur2013nonlinear, servidio2017magnetospheric, adkins2018solvable, pezzi2018velocity,zhdankin2022generalized,zhdankin2022non}, and is closely related to the part of the traditional entropy, $S = -\iint dx dv \, f \log f$, associated with the perturbed distribution function and entering additively in the free-energy invariant of $\delta \! f$ gyrokinetics \cite{krommes1994role,hallatschek2004thermodynamic,schekochihin2008gyrokinetic,schekochihin2009astrophysical,abel2013multiscale}. The conservation of $C_2$ is broken by particle collisions, and, in particular, when collisions are modeled as a linear diffusion operator in velocity space, as they are in this paper, the collisional dissipation of $C_2$ is negative-definite. Because the time irreversibility of our system can be tracked via non-conservation of $C_2$, it can be used as a generalized (negative) entropy of the distribution function.

The diffusion of $\langle f \rangle$ by the electric field has the side effect of injecting $\delta C_2 = (1/L) \iint dx dv \, \langle \delta \! f^2 \rangle/2$ fluctuations at large scales. These are then cascaded to small scales in both position and velocity space, where they are ultimately dissipated by collisions. This phase-space cascade of $\delta C_2$ is due to both linear phase mixing and nonlinear interactions between particles and the stochastic electric field. We analyze this cascade by computing the steady-state Fourier spectrum and fluxes of $\delta C_2$ in both position and velocity space, with dual variables (wavenumbers) $k$ and $s$, respectively.

In the absence of nonlinearity, linear phase mixing advects the spectrum from low to high $|s|$, giving rise to an `inertial range' in $s$ where there is a constant flux of $\delta C_2$ from injection to dissipation scales, at every $k$ \cite{zocco2011reduced,kanekar2015fluctuation}. The resulting steady-state spectrum is flat in the inertial range, with an exponential cutoff at a $\nu$-dependent collisional scale $s_{\nu}$.

Under the Kraichnan model, we find that the nonlinear term in the evolution equation for the Fourier spectrum becomes a fractional Laplacian operator \cite{2012hitchhiker,pozrikidis2018fractional,lischke2020fractional} in $k$ space, which leads to anomalous diffusion \cite{bouchaud1990anomalous,zaslavsky2008chaos}. Whereas fractional Laplacians (and fractional derivatives in general) are usually introduced in an ad-hoc manner to model systems with anomalous diffusion, both in plasma physics \cite{del2004fractional,del2005nondiffusive,del2006fractional} and a wide variety of other contexts \cite{hilfer2000applications}, here it emerges naturally as a result of our assumptions about the electric field.

The linear phase mixing and the turbulent fractional diffusion intertwine the position- and velocity-space cascades in such a way that the resulting spectrum is a self-similar function in the $(k,s)$ plane, with power-law asymptotics at large $k$ and $s$. Even though the Kraichnan model is fully analytically solvable, we can also recover these asymptotic scalings via a phenomenological theory whose key assumption is that the cascade is governed by a `critical balance' in phase space between the linear and nonlinear time scales.

The $\delta C_2$ flux has components in both $k$ and $s$ directions. The flow of $\delta C_2$ occurs along outward unwinding spirals in $(k,s)$ space. This circuitous route to dissipation scales is due to the nonlinearity generating modes that can linearly propagate from high to low $|s|$, called `phase-unmixing' modes \cite{schekochihin2016phase}, a stochastic generalization of the textbook phenomenon of plasma echo \cite{gould1967plasma,malmberg1968plasma}. The net result after adding together contributions to the flux from the phase-mixing and phase-unmixing modes is that $\delta C_2$ is cascaded to both high $s$ and high $k$ simultaneously, and collisional dissipation only occurs at scales comparable to, or smaller than, the dissipation length and velocity scales, both of which tend to zero when the collision frequency $\nu$ does. Integrating over velocity (spatial) wavenumbers, the flux of $\delta C_2$ in $k$ ($s$) space is constant down to these dissipation scales, beyond which perturbations are thermalized by collisions. The rate of collisional dissipation is finite and independent of the collision frequency as $\nu \rightarrow 0^{+}$—a clear analytical example of a `dissipative anomaly' in a kinetic system. This turbulent dissipation ultimately mediates the irreversibility of the stochastic heating. 

The rest of this paper is organized as follows. In Section \ref{Kraichnan}, we introduce the Vlasov-Kraichnan model. In Section \ref{twiddle_theory}, we construct a phenomenological theory that captures the asymptotic scalings of the Fourier spectrum of $\delta C_2$. Then, in Section \ref{forced}, we directly calculate the 1-D phase-space fluxes of $\delta C_2$ in Fourier space in a statistical steady state. In Section \ref{spec_self_similar}, we calculate the inertial-range Fourier spectrum and its corresponding 2-D fluxes in $(k,s)$ space. Finally, in Section \ref{discussion}, we conclude our results and discuss their implications. Supplementary calculations and discussions are exiled to Appendices \ref{heating_sec}-\ref{phases_k_flux}.

\section{Kraichnan model for a 1D-1V electrostatic plasma} \label{Kraichnan}
We consider a test-particle species composed of particles with charge $q$ and mass $m$ in a 1D periodic box of length $L$, and subject to an external electric field $E$. At $t = 0$, we assume the particle distribution function $f$ to have no spatial variation, but we keep its velocity dependence generic, only assuming that $f$ is square-integrable and has finite kinetic energy. We denote the number density of the distribution function as $n_0$ and the initial thermal velocity as $v_{\mathrm{th},0} = \sqrt{2T_0/m}$, where $T_0$ is the initial temperature of the particles. An example initial condition with these properties is a Maxwellian,
\begin{equation} \label{initial_condition}
    f(x,v,t = 0) = \frac{n_0}{\sqrt{\pi} v_{\mathrm{th},0}}e^{-v^2/v_{\mathrm{th},0}^2} \equiv F_M(v).
\end{equation}

The Vlasov equation for the particle distribution function is
\begin{equation} \label{Vlasov}
    \frac{\partial f}{\partial t} + v \frac{\partial f}{\partial x} + E \, \frac{\partial f}{\partial v} = C[f],
\end{equation}
where $C[f]$ is the collision operator, and we have absorbed $q$ and $m$ into the definition of $E$, denoting $q E /m \rightarrow E$.

We assume $E$ to be a Gaussian white-noise field, with zero mean and correlation function
\begin{equation} \label{electric_field_correlator}
    \left\langle E(x, t) E(x', t') \right\rangle = 2 \, D(x,x') \, \delta(t-t'),
\end{equation}
where $\langle ... \rangle$ denotes ensemble averaging over realizations of $E$ and $\delta$ is a Dirac delta distribution. We assume $E$ to be statistically homogeneous and isotropic in space, so $D(x,x') = D(r)$, where~${r = |x-x'|}$. We choose
\begin{equation} \label{Dr}
    D(r) = \sum_k e^{i k r} \hat{D}(k),
\end{equation}
where $k \in (2\pi/L) \mathbb{Z}$ and
\begin{equation} \label{hatD}
    \hat{D}(k) = D \, \frac{e^{-(\eta k)^2}}{(k^2+L_{E}^{-2})^{(\alpha+1)/2}}.
\end{equation}
Here, $D$ is a constant diffusion coefficient with dimensions (length$^{1-\alpha}$) $\times$ (time$^{-3}$), $ 0 < \alpha \leq 2$, and $L_{E}$ and $\eta$ represent the integral length scale and dissipation scale, respectively, of the stochastic electric field.

The electric field is chosen so that its correlation function \eqref{electric_field_correlator} has a power-law spectrum $ \propto |k|^{-(\alpha+1)}$ in the inertial range $1/L_{E} \ll k \ll 1/\eta$; note that \eqref{hatD} is defined in a similar way to the velocity field in the fluid passive scalar Kraichnan model \cite{eyink2000self,falkovich2001particles}. We identify two distinct regimes: $\alpha < 2$ and $\alpha = 2$. When $\alpha < 2$, the field is multiscale, reminiscent of turbulent fields in fully developed turbulence. When $\alpha = 2$, the spectrum is sufficiently steep that the field is effectively single-scale. This case is known as the Batchelor regime \cite{batchelor1959small}. While there are important differences between the two regimes \cite{falkovich2001particles}, some of which we will discuss, many of the properties of the model considered in this paper will be qualitatively the same in both regimes.

It will be useful to decompose the distribution function into its mean and fluctuating parts:
\begin{equation} \label{f_decomp}
    f = \langle f \rangle + \delta \! f.
\end{equation}
We make no assumption of $\delta \! f$ being small compared to~$\langle f \rangle$.

The effect of collisions will be to wipe out fine-scale velocity-space structure in the distribution function. To model this in the simplest possible way, we ignore collisions between our test-particle species and the other species in the plasma, and represent collisions within the test species as a linear diffusion in velocity space acting only on $\delta \! f$, viz.,
\begin{equation} \label{approx_coll_op_delta_f}
  C[\delta \! f] = \nu \frac{\partial^2 \delta \! f}{\partial v^2},
\end{equation}
where $\nu$ is the collision frequency (multiplied by $v_{\mathrm{th},0}^2$), which we consider to be vanishingly small, taking $\nu \rightarrow 0^{+}$. It is not a problem that \eqref{approx_coll_op_delta_f} does not conserve energy or vanish on a Maxwellian because collisions will only matter for the parts of $\delta \! f$ with sharp gradients in $v$ \footnote{This approximation may seem overly simplified; for example, it neglects collisions acting on $\langle f \rangle$, collisional drag, and nonlinearity \cite{helander2005collisional}. However, we expect these effects to be subdominant compared to collisional diffusion of $\delta \! f$ for the following reasons. In Section \ref{SA}, we find that $\langle f \rangle$ satisfies a diffusion equation in velocity space. Its gradients therefore become ever smaller in time, so weak collisions will be negligible for $\langle f \rangle$. Collisional drag should be negligible in the limit $\nu \rightarrow 0^{+}$ because it contains one fewer velocity derivative than collisional diffusion. We assume that dropping nonlinearity in the collision operator is reasonable because we anticipate that $\delta \! f$ will be small at the scales where collisions matter. Indeed, we find in Sections \ref{twiddle_theory} and \ref{spec_self_similar} that the Fourier spectrum of $\delta \! f$ decays as a power law in wavenumber space, with collisional cutoffs occurring at large wavenumbers, where the spectrum is small relative to its magnitude at large scales.}.

\subsection{Stochastic heating} \label{SA}

We first work out the effect of the turbulent electric field on the mean distribution function. Ensemble averaging \eqref{Vlasov} over realizations of the stochastic electric field, we get

\begin{equation} \label{Vlasov_avg}
    \frac{\partial \langle f \rangle }{\partial t} + v \frac{\partial \langle f \rangle }{\partial x} +  \left\langle E \frac{\partial \delta \! f}{\partial v} \right\rangle = 0.
\end{equation}
The equation for $\delta \! f$ is then
\begin{align} \label{delta_f_eq}
    \frac{\partial \delta \! f}{\partial t} + v \frac{\partial \delta \! f}{\partial x} + E \frac{\partial \delta \! f}{\partial v}\nonumber - &\left\langle E \frac{\partial \delta \! f}{\partial v} \right\rangle \\ =& - E \frac{\partial \langle f \rangle}{\partial v} + \nu \frac{\partial^2 \delta \! f}{\partial v^2}.
\end{align}
To compute the ensemble average of the nonlinear term in \eqref{Vlasov_avg}, we apply the Furutsu-Novikov theorem \cite{furutsu1964statistical,novikov1965functionals} for splitting correlators. For a Gaussian field $E$ that depends on variables $\mathbf{q}$, this theorem states that
\begin{equation} \label {NF}
    \langle E(\mathbf{q}) F\left[E\right] \rangle = \int d\mathbf{q'} \left \langle E(\mathbf{q}) E(\mathbf{q'}) \right \rangle \left \langle \frac{\delta F\left[E\right]}{\delta E(\mathbf{q'})}  \right \rangle,
\end{equation}
where $F\left[E\right]$ is a differentiable functional of $E$. Formally integrating \eqref{delta_f_eq} with respect to time, we have that
\begin{align} \label{delta_f_integrate}
    \delta \! f \nonumber  = - \int^{t} dt'' \bigg[v \frac{\partial \delta \! f}{\partial x} & + E \frac{\partial \delta \! f}{\partial v} -\left\langle E \frac{\partial \delta \! f}{\partial v} \right\rangle \\ & + E \frac{\partial \langle f \rangle}{\partial v} - \nu \frac{\partial^2 \delta \! f}{\partial v^2} \bigg](t'').
\end{align}
Combining \eqref{NF} and \eqref{delta_f_integrate}, we have
\begin{align}
    &\left\langle E(x,t) \delta \! f(x,t) \right\rangle \nonumber \\ &= \int dt' \int dx' \left\langle E(x,t) E(x',t') \right\rangle \left \langle \frac{\delta \left[ \delta \! f(x,t) \right]}{\delta E(x',t')} \right \rangle \nonumber \\ &= - \int dx' D(x-x') \delta(x-x') \frac{\partial \langle f \rangle}{\partial v} = -D(0) \frac{\partial \langle f \rangle}{\partial v}.
\end{align}
Therefore, \eqref{Vlasov_avg} becomes 
\begin{equation} \label{vlasov_avg_diff}
    \frac{\partial \langle f \rangle }{\partial t} = D_0\frac{\partial^2 \langle f \rangle}{\partial v^2},
\end{equation}
where $D_0 = D(0)$ is the `turbulent collisionality.' We have dropped the streaming term in \eqref{Vlasov_avg} because our initial condition is spatially homogeneous, so $\langle f \rangle$ at future times does not depend on $x$. The solution to \eqref{vlasov_avg_diff} is
\begin{align} \label{f_avg_sol}
   \langle f \rangle = \int dv' \, f(v',t=0) \frac{1}{\sqrt{4 \pi D_0t}}e^{-(v-v')^2/4D_0t}.
\end{align}
The mean kinetic-energy density $\langle K \rangle$ of this distribution function is
\begin{align} \label{KE_evol}
     \langle K \rangle \equiv \int dv \frac{m v^2}{2} \langle f \rangle = K_0 + m n_0 D_0 t,
\end{align}
where $K_0$ is the initial kinetic-energy density. For the Maxwellian initial condition \eqref{initial_condition}, \eqref{f_avg_sol} becomes simply
\begin{equation} \label{f0_time_max}
    \langle f \rangle = \frac{n_0}{\sqrt{\pi} v_{\mathrm{th}}}e^{-v^2/v_{\mathrm{th}}^2},
\end{equation}
with a growing thermal speed:
\begin{equation} \label{vth}
    v_{\mathrm{th}} = \sqrt{v_{\mathrm{th},0}^2 + 4 D_0 t}.
\end{equation}

As particles get stochastically accelerated by the electric field, they undergo Brownian random walks in velocity space, leading to bulk heating of the distribution function \cite{sturrock1966stochastic}, with secularly growing kinetic energy \eqref{KE_evol}. In magnetized plasmas, this phenomenon is usually referred to as stochastic heating \cite{chandran2010perpendicular,verscharen2019multi,cerri2021stochastic}.

\subsection{\texorpdfstring{$C_2$}{C2} budget: injection and dissipation} \label{C2_balance}
Because the stochastic electric field continuously heats the distribution function, viz., \eqref{KE_evol}, energy is not a conserved quantity in our system. However, what is conserved in the absence of collisions is the quadratic quantity
\begin{equation} \label{C_2}
    C_2 = \frac{1}{L} \iint dx \, dv \, \frac{1}{2} \langle f^2 \rangle = C_{2, 0} + \delta C_2,
\end{equation}
where
\begin{equation}
    C_{2, 0} = \frac{1}{L} \iint dx \, dv \, \frac{1}{2} \langle f \rangle ^2,
\end{equation}
\begin{equation} \label{deltaC2}
    \delta C_2 = \frac{1}{L} \iint dx \, dv \, \frac{1}{2} \langle \delta \! f^2 \rangle.
\end{equation}
$C_2$ can only change via collisions. Using \eqref{Vlasov} and \eqref{approx_coll_op_delta_f}, we have
\begin{align} \label{dC2dt}
    \frac{d C_2}{d t} &= \frac{1}{L} \iint dx \, dv \, \langle \delta \! f \, C[\delta \! f] \rangle \nonumber \\ &= - \frac{\nu}{L} \iint dx \, dv \, \left \langle \left |\frac{\partial \delta \! f}{\partial v} \right|^2 \right\rangle.
\end{align}
Thus, when collisions are approximated as a linear diffusion in velocity space, they provide negative-definite dissipation of $C_2$. Because $-C_2$ is conserved in the absence of collisions and positive-definitely increased by collisions, we can interpret it as a `generalized entropy' of the distribution function. We will henceforth refer to $-C_2$ and entropy synonymously.

Stochastic heating, which is a collisionless process, is accompanied by the decrease of $C_{2,0}$. Indeed, using \eqref{vlasov_avg_diff} and integrating by parts, we have
\begin{align} \label{dC20dt}
     \frac{d C_{2,0}}{d t} &= - \frac{D_0}{L} \iint dx dv \,  \left|\frac{\partial \langle f \rangle }{\partial v} \right|^2 \nonumber \\ &= -D_0 \, \int dv \,  \left|\frac{\partial \langle f \rangle }{\partial v} \right|^2 \leq 0
\end{align}
generically, and for a Maxwellian in particular,
\begin{equation} \label{dC20dt_Maxwellian}
    \frac{d C_{2,0}}{d t} = -\frac{n_0^2 \sqrt{m}}{8 \sqrt{\pi}} \frac{1}{T^{3/2}} \frac{d T}{dt},
\end{equation}
where $T = m v_{\mathrm{th}}^2/2$, with $v_{\mathrm{th}}$ given by \eqref{vth}. To work out the $\delta C_2$ budget, we can combine \eqref{C_2} and \eqref{dC2dt}, giving
\begin{equation} \label{deltaC2balance}
    \frac{d \delta C_2}{d t} = D_0 \, \int dv \,  \left|\frac{\partial \langle f \rangle }{\partial v} \right|^2 - \frac{\nu}{L} \iint dx \, dv \, \left \langle \left |\frac{\partial \delta \! f}{\partial v} \right|^2 \right\rangle.
\end{equation}
Thus, without collisions, as $C_{2, 0}$ decreases as a result of the stochastic heating of $\langle f \rangle$, $\delta C_2$ increases to maintain entropy balance. Once $\delta \! f$ has developed sharp enough gradients, collisions dissipate $\delta \! f$, increasing the total entropy. The irreversiblity of stochastic heating therefore hinges on the collisional dissipation of $\delta \! f$.

If the $\delta C_2$ fluctuations evolve faster than the mean $C_{2,0}$ and reach a quasi-steady state (as we shall argue that they do), then \eqref{deltaC2balance} becomes
\begin{equation} \label{deltaC2balance_ss}
    D_0 \int dv \,  \left|\frac{\partial \langle f \rangle }{\partial v} \right|^2 = \frac{\nu}{L} \iint dx \, dv \, \left \langle \left |\frac{\partial \delta \! f}{\partial v} \right|^2 \right\rangle.
\end{equation}
In the case where the initial condition is Maxwellian, combining \eqref{dC20dt_Maxwellian} and \eqref{dC20dt} and substituting this expression into \eqref{deltaC2balance_ss} yields a direct balance between the heating rate of $\langle f \rangle$ and the collisional dissipation of $\delta \! f$. 

The velocity-space gradients of $\langle f \rangle$ inject $\delta C_2$ at large scales, and collisions dissipate $\delta C_2$ at small scales. As in any prototypical turbulent system, the steady-state balance \eqref{deltaC2balance_ss} between injection and dissipation at such disparate scales can hold if there is a constant-flux cascade bridging them. This is precisely what we will find in the following sections, viz., a phase-space cascade of $\delta C_2$ in both position and velocity space. This cascade will be our focus for the rest of this paper.

Before continuing, we note that in Appendix \ref{heating_sec}, we discuss alternative formulations of the thermodynamics of our system in terms of other collisionless invariants, and why in this work we have chosen to study $C_2$ over those other invariants.

\subsection{Evolution of Fourier spectrum} \label{Fourier_eq}
We analyze how $\delta C_2$ is partitioned between scales in phase space via its Fourier spectrum. We define
\begin{equation} \label{fourier_def}
    \delta \hat{f}(k,s) = \frac{1}{2 \pi L} \iint dx \, dv \, e^{-i(kx-sv)} \, \delta \! f(x,v),
\end{equation}
\begin{equation} \label{fourier_def_f_avg}
    \hat{\langle f \rangle}(s) = \frac{1}{2 \pi} \int dv \, e^{isv} \, \langle f \rangle(v),
\end{equation}
and
\begin{equation} \label{fourier_def_elc}
    \hat{E}(k) = \frac{1}{L} \int dx \, e^{-ikx} \, E(x),
\end{equation}
where $k$ and $s$ are dual variables to $x$ and $v$, respectively. Fourier transforming \eqref{delta_f_eq}, we get that $\delta \hat{f}(k,s)$ satisfies
\begin{align} \label{fhat_eq}
    \frac{\partial \delta \hat{f}}{\partial t} &+ k\frac{\partial \delta \hat{f}}{\partial s} \nonumber \\ &- i s \,  \sum_p \, \left[\hat{E}(p) \delta \hat{f}(k-p) -\left\langle \hat{E}(p) \delta \hat{f}(k-p) \right\rangle \right] \nonumber \\ &= i s \, \hat{E} \hat{\langle f \rangle} -\nu s^2 \delta \hat{f}.
\end{align}
We define the Fourier spectrum as
\begin{equation}
    \hat{F}(k,s) = \frac{1}{2} \langle |\delta \hat{f}(k,s)|^2 \rangle,
\end{equation}
which satisfies Parseval's theorem,
\begin{equation} \label{Planch}
    \delta C_2 = \frac{1}{L} \iint dx \, dv \, \frac{1}{2} \langle \delta \! f^2 \rangle = 2 \pi \sum_k \, \int ds \, \hat{F}(k,s),
\end{equation}
has the budget equation (equivalent to \eqref{deltaC2balance} in spectral space)
\begin{equation} \label{C2_Fhat_balance}
    \frac{d \delta C_2}{d t} = 2 \pi \left[\sum_k \, \int ds \, \hat{S} - 2 \nu \sum_k \, \int ds \, s^2 \hat{F} \right],
\end{equation}
and satisfies the evolution equation
\begin{equation} \label{spec_eq_FL_noflux}
    \frac{\partial \hat{F}}{\partial t} + k\frac{\partial \hat{F}}{\partial s} + \kappa \, s^2 (-\Delta_k)^{\alpha/2}\hat{F} = \hat{S} - 2 \nu s^2 \hat{F}.
\end{equation}
Equation \eqref{spec_eq_FL_noflux} is important, and its analysis will be the focus of the rest of this paper. There are several steps required to derive \eqref{C2_Fhat_balance} and \eqref{spec_eq_FL_noflux} from \eqref{fhat_eq},  primarily using~\eqref{NF} to calculate correlation functions involving the electric field. These steps are technical, so we detail them in Appendix \ref{error_estimate_sec}.

The source term in \eqref{C2_Fhat_balance} and \eqref{spec_eq_FL_noflux} is
\begin{equation} \label{source_sk}
    \hat{S}(k,s) = \hat{D}(k) s^2 \langle \hat{f} \rangle^2,
\end{equation}
which for a Maxwellian initial condition evaluates to
\begin{equation} \label{source_sk_Maxwellian}
    \hat{S}(k,s) = \frac{n_0^2}{(2 \pi)^2} \, \hat{D}(k) \, s^2 \, e^{-s^2v_{\mathrm{th}}^2/2},
\end{equation}
where $v_{\mathrm{th}}$ is given by \eqref{vth}.

In \eqref{spec_eq_FL_noflux}, $(-\Delta_k)^{\alpha/2}$ is a fractional Laplacian \cite{2012hitchhiker,pozrikidis2018fractional,lischke2020fractional} of order $\alpha/2$ (in $k$ space), and $\kappa$ is a turbulent diffusion coefficient $\propto L D$, for which the exact expression is given in Appendix \ref{error_estimate_sec}. To obtain this term, we have taken the limit $\eta \rightarrow 0^{+}$ in \eqref{hatD}, which is analogous to taking the zero-viscosity limit in hydrodynamic turbulence, and $k L_{E} \gg 1 $, which is a convenient limit to take to focus on how the distribution function is stirred in the `inertial range' of the stochastic electric field.

The fractional Laplacian is a non-local integral operator, generalizing the Laplacian to non-integer order. Its Fourier transform satisfies
\begin{equation} \label{Fourier_Transform_fracLap}
\frac{L}{2 \pi} \int dk \, e^{ikr} (-\Delta_k)^{\alpha/2}\hat{F}(k,s) = |r|^{\alpha} F(r,s),
\end{equation}
where
\begin{equation} \label{Fhat_r}
    {F}(r,s) = \frac{L}{2 \pi} \int dk \, e^{ikr} \hat{F}(k,s).
\end{equation}
Note that we have converted sums over $k$ into integrals (multiplied by the inverse step size $L/2 \pi$).

Mathematically, the fractional Laplacian describes abstract `particles' with coordinates $(k,s)$ undergoing random jumps in $k$ space, so-called Lévy flights, which leads to superdiffusion \cite{bouchaud1990anomalous,zaslavsky2008chaos}. In the limit $\alpha \rightarrow 2^{-}$, the fractional Laplacian becomes (minus) a regular Laplacian \cite{2012hitchhiker}, corresponding to regular Brownian motion. This limit, the Batchelor regime, was studied before in \cite{adkins2018solvable}, although we shall present some new conclusions about it below.

In the following sections, we will solve for steady-state solutions of \eqref{spec_eq_FL_noflux}. To this end, we further assume that the source is localized at small $(k,s)$ and injects $\delta C_2$ at a constant rate,
\begin{equation} \label{const_inj}
   L \iint dk \, ds \, \hat{S}(k,s) = \varepsilon > 0.
\end{equation}
These assumptions may appear questionable in view of \eqref{deltaC2balance} and \eqref{dC20dt} (is the source really constant in time?) and of \eqref{source_sk} and \eqref{hatD} (is it really local?). Before continuing, we address these concerns.

Regarding the constancy of the source, we observe that in~\eqref{deltaC2balance}, the injection rate of $\delta C_2$ via the stochastic heating of $\langle f \rangle$, using \eqref{dC20dt} and dimensional analysis, scales as
\begin{equation} \label{source_C2_time_generic}
    \varepsilon = -\frac{d C_{2,0}}{dt} \propto \frac{D_0 n_0^2}{v_{\mathrm{th}}^3} = \frac{D_0 n_0^2}{(v_{\mathrm{th},0}^2 + 4 D_0 t)^{3/2}}.
\end{equation}
This scaling is particularly obvious, from \eqref{dC20dt_Maxwellian}, for the case of a Maxwellian initial condition. Comparing \eqref{source_C2_time_generic} and \eqref{const_inj}, we see that the injection rate is, in fact, time-dependent, $\propto t^{-3/2}$, even though in our calculations we would like to treat $\varepsilon$ as a constant. However, if the cascade of $\delta C_2$ is set up quickly compared to the decay of the source (which, as will be discussed in Section \ref{C2_capacity}, it is on a $\nu$-independent time scale when $\alpha < 2$ and a time scale $\propto | \log \nu |$ when $\alpha = 2$), it is reasonable to treat the source as approximately constant on the time scale of the $\delta \! f$ evolution, and then solve for the steady-state spectrum. This becomes an ever-better approximation at long times, because the rate of change of \eqref{source_C2_time_generic} is a decreasing function of time. Also note that, as long as $\varepsilon > 0$, its actual value is not important; because the spectrum-evolution equation \eqref{spec_eq_conv} is linear, it has no special amplitude scale, and so the size of $\varepsilon$ is just a global modifier of the spectrum's amplitude.

Regarding the locality of the source, it is easily satisfied in $s$ space, as the velocity derivatives of~$\langle f \rangle$ become ever smaller over time as the distribution function stochastically heats. This is clearly seen in the case of a Maxwellian initial condition, where the source \eqref{source_sk_Maxwellian} is a Gaussian in $s$, with a width $\sim 1/v_{\mathrm{th}}$. In $k$ space, the source is not, in fact, truly localized, since $\hat{D}(k)$ is a power law, viz., \eqref{hatD}, so fluctuations are injected at all scales where the electric field exists. However, this turns out not to be fatal to our formalism: we will neglect the source in our solving for the inertial-range spectrum in Sections \ref{twiddle_theory} and~\ref{spec_self_similar} and argue \textit{a posteriori} in Section \ref{PD_section} that this choice was justified.

\section{Phenomenological theory of phase-space cascade} \label{twiddle_theory}

In Sections \ref{forced} and \ref{spec_self_similar}, we perform a detailed analysis of \eqref{spec_eq_FL_noflux}. However, the key results of those sections can, in fact, be intuited via a much simpler route, which we follow here first.

As discussed in Section \ref{C2_balance}, stochastic heating can only truly be made irreversible by particle collisions. For collisions to be relevant even in the limit $\nu \rightarrow 0^{+}$, we conjecture that $\delta C_2$ undergoes a cascade in both position and velocity space. Analogously to phenomenological theories of hydrodynamic turbulence \cite{kolmogorov1941b,frisch1995turbulence}, we assume that there exists an inertial range in phase space, whereby the flux $\varepsilon$ is processed between scales, from the injection to dissipation range. Under this assumption, our task is then to ascertain the form of the spectrum $\hat{F}$ in $(k,s)$ space in the inertial range.

Fluctuations of $\delta \! f$ develop fine-scale structure in velocity space via linear phase mixing, which  manifests in~\eqref{spec_eq_FL_noflux} as advection of $\hat{F}$ in $s$ at the rate $k$. Dimensionally, the phase-mixing time is, therefore,
\begin{equation} \label{p_time}
    \tau_p \sim \frac{s}{k}.
\end{equation}
Likewise, $\delta \! f$ develops fine-scale structure in position space via nonlinear mixing by the stochastic electric field, which manifests in \eqref{spec_eq_FL_noflux} as fractional diffusion of $\hat{F}$ in $k$ with the diffusion coefficient $\kappa s^2$. Dimensionally, using~\eqref{Fourier_Transform_fracLap}, the turbulent-diffusion time is, therefore, 
\begin{equation} \label{d_time}
    \tau_d \sim \frac{k^{\alpha}}{\kappa s^2}.
\end{equation}
The ratio of these time scales (taken to the power ${1/(\alpha+1)}$ for analytical convenience and in anticipation of the results of Section \ref{spec_self_similar}) is
\begin{equation} \label{ratio_time_scale}
    \xi =\left(\frac{\tau_d}{\tau_p}\right)^{1/(\alpha+1)} = \frac{k}{(\kappa \, s^3)^{1/(\alpha+1)}}.
\end{equation}
We conjecture that the structure of $\hat{F}$ in $(k,s)$ space is governed by the parameter $\xi$. In particular, we assume that the spectrum is a product of power laws in $k$ and $s$, with different scaling exponents depending on whether $\xi$ is small or large:

\begin{equation} \label{spec_unknown_exp}
    \hat{F}(k,s) \propto
\begin{cases} 
     k^{a} s^{-b}, \qquad \qquad & \xi \ll 1, \\
     k^{-c} s^{d}, \qquad \qquad & \xi \gg 1.
   \end{cases}
\end{equation}

When $\xi \ll 1$, the turbulent-diffusion time is much shorter than the phase-mixing time, so we expect the inertial-range spectrum to satisfy, to lowest order in $\xi$,
\begin{equation} \label{homog_frac}
     \kappa s^2 \,(-\Delta_k)^{\alpha/2}\hat{F} = 0,
\end{equation}
and therefore, to be independent of $k$. Consequently, $a = 0$ in \eqref{spec_unknown_exp}.

When $\xi \gg 1$, the phase-mixing time is much shorter than the turbulent-diffusion time. Then, to lowest order in $1/\xi$, we expect the spectrum to satisfy
\begin{equation} \label{linear_constant_flux}
     k\frac{\partial \hat{F}}{\partial s} = 0,
\end{equation}
and, therefore, to be independent of $s$, the same as in the linear regime \cite{kanekar2015fluctuation,zocco2011reduced}. Consequently, $d = 0$ in \eqref{spec_unknown_exp}.

To find $b$ and $c$, we invoke our initial assumption of a constant-flux cascade. As in any Kolmogorov-style theory, we need a prescription for the cascade time $\tau_{c}$, which is the typical time for $\delta C_2$ to be transferred across phase-space scales $(\ell, u)$, where $\ell \sim 1/k$ and $u \sim 1/s$. We conjecture that the cascade time is set by the phase-mixing and turbulent-diffusion time scales \eqref{p_time} and \eqref{d_time}, and that the latter two must balance along the path of the cascade:
\begin{align} \label{tau_c}
&\tau_{\mathrm{c}} \sim \tau_{\mathrm{p}} \sim \tau_{\mathrm{d}} \nonumber \\ &\implies \tau_{\mathrm{c}} \sim \kappa^{-1/3} \ell^{(2-\alpha)/3} \sim \kappa^{-1/(\alpha + 1)} u^{(2-\alpha)/(\alpha + 1)}.
\end{align}
In wavenumber space, this condition is satisfied when $\xi \sim 1$, i.e.,
\begin{equation} \label{CB}
    s \sim \kappa^{-1/3} k^{(\alpha+1)/3}.
\end{equation}
This is a kinetic, phase-space analogue of the critical-balance conjecture in magnetohydrodynamic turbulence \cite{goldreich1995toward,schekochihin2022mhd}.

Assuming a constant flux of $\delta C_2$ in position space, using the $\ell$ scaling in \eqref{tau_c}, and using \eqref{Planch} to relate fluctuations in real space and wavenumber space, we get
\begin{align} \label{spec_1D_k_twiddle}
     \frac{v_{\mathrm{th}} \delta \! f_{\ell}^2}{\tau_{c}}  \sim \varepsilon & \implies v_{\mathrm{th}} \delta \! f_{\ell}^2 \sim \varepsilon \, \kappa^{-1/3} \ell^{(2-\alpha)/3} \nonumber \\  &  \iff L \int ds \hat{F} \sim \varepsilon \, \kappa^{-1/3}  k^{-(5-\alpha)/3},
\end{align}
where $\delta \! f_{\ell} $ is the characteristic amplitude of $\delta \! f$ at spatial scale $\ell$. Let us compare this result to the 1-D $k$ spectrum that follows from \eqref{spec_unknown_exp}. We assume (and verify \textit{a posteriori}) that $b > 1$, so that the integral of $\hat{F}$ over $s$ is dominated by the region $\xi \gtrsim 1$, i.e.,
\begin{equation} \label{s_less_CB}
    s \lesssim \kappa^{-1/3} k^{(\alpha+1)/3}.
\end{equation}
This gives
\begin{equation} \label{spec_1D_k_integral_twiddle}
    \int ds \, \hat{F} \propto \int^{\kappa^{-1/3} k^{(\alpha+1)/3}}_{0} ds \, k^{-c} \propto k^{-c + (\alpha+1)/3}.
\end{equation}
Requiring consistency between \eqref{spec_1D_k_integral_twiddle} and \eqref{spec_1D_k_twiddle} yields $c = 2$.

We can now perform the same exercise in velocity space, viz.,
\begin{align} \label{spec_1D_s_twiddle}
     \frac{v_{\mathrm{th}} \delta \! f_{u}^2}{\tau_{c}} \sim \varepsilon & \implies v_{\mathrm{th}} \delta \! f_{u}^2 \sim \varepsilon \kappa^{-1/(\alpha + 1)} u^{(2-\alpha)/(\alpha + 1)} \nonumber \\  &  \iff L \int dk \hat{F} \sim \varepsilon \, \kappa^{-1/(\alpha + 1)} s^{-3/(\alpha + 1)},
\end{align}
where $\delta \! f_{u} $ is the characteristic amplitude of $\delta \! f$ at velocity scale $u$. On the other hand, using \eqref{spec_unknown_exp} and assuming that the integral over $k$ of $\hat{F}$ is dominated by the region $\xi \lesssim 1$, or
\begin{equation}
    k \lesssim \kappa^{1/(\alpha+1)} s^{3/(\alpha+1)},
\end{equation}
we have
\begin{equation} \label{spec_1D_s_integral_twiddle}
     \int dk \, \hat{F} \propto \int^{\kappa^{1/(\alpha+1)} s^{3/(\alpha+1)}}_{0} dk \, s^{-b} \propto s^{-b + 3/(\alpha + 1)}.
\end{equation}
Requiring consistency between \eqref{spec_1D_s_integral_twiddle} and \eqref{spec_1D_s_twiddle} yields $b = 6/(\alpha+1)$.

\begin{figure}
	\centering
	\includegraphics[width=1\linewidth]{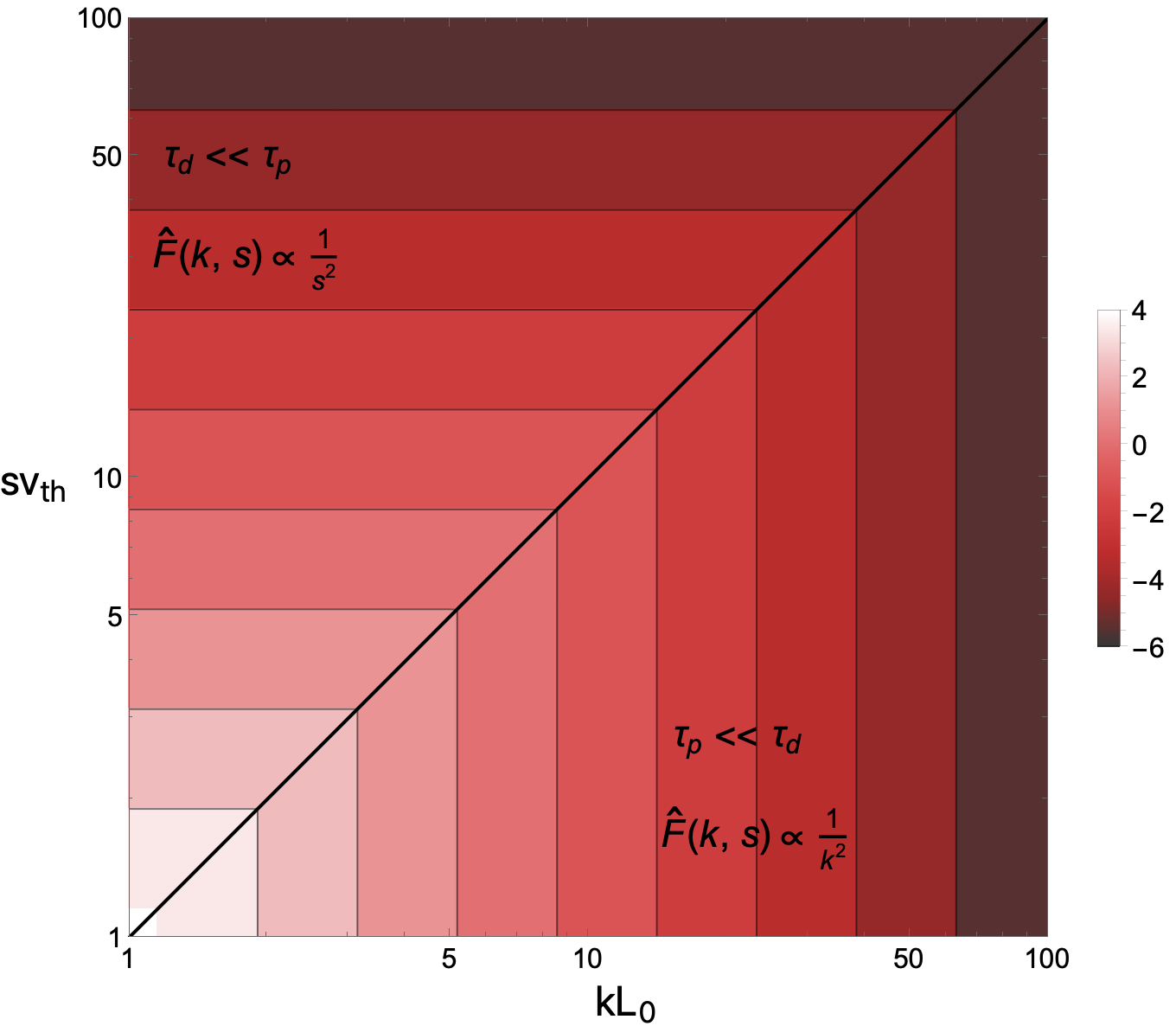}
\caption{\label{fig:spec_cartoon_a2} Cartoon contour plot of $\log L_0^2 \, v_{\mathrm{th}}^{6/(\alpha + 1)} \hat{F}(k L_0 , s v_{\mathrm{th}})$ vs. $(k L_0 ,s v_{\mathrm{th}})$, for $\alpha = 2$, using the piecewise scalings \eqref{asymptotics_spec_twiddle}. We use normalized units $s v_{\mathrm{th}}$ and $k L_0$, where $L_0$ is defined in \eqref{L0} (see the discussion at the beginning of Section \ref{PD_section}). The black line is the critical-balance line~\eqref{CB}, $s v_{\mathrm{th}} \sim k L_0$. In the $\tau_p \ll \tau_d$ ($\xi \gg 1$) region, the spectrum scales as $k^{-2}$, and in the $\tau_d \ll \tau_p$ ($\xi \ll 1$) region, the spectrum scales as $s^{-2}$.}
\end{figure}

Assembling the scalings that we have surmised above, we find that the spectrum \eqref{spec_unknown_exp} is
\begin{equation} \label{asymptotics_spec_twiddle}
    \hat{F}(k,s) \sim \varepsilon L^{-1}
\begin{cases} 
      \kappa^{-2/(\alpha+1)} \, s^{-6/(\alpha + 1)}, \quad \qquad & \xi \ll 1, \\
      k^{-2}, \quad \qquad & \xi \gg 1.
   \end{cases}
\end{equation}
For $\alpha = 2$, this was derived, by means of a formal solution, in \cite{adkins2018solvable}, but the phenomenological argument and physical interpretation presented here are new. Note that the dimensional factors in \eqref{asymptotics_spec_twiddle} come from using Parseval's theorem \eqref{Planch} together with demanding that the integrals of the spectrum in \eqref{spec_1D_k_integral_twiddle} and \eqref{spec_1D_s_integral_twiddle} satisfy the constant-flux relations in position and velocity space, respectively, viz., \eqref{spec_1D_k_twiddle} and \eqref{spec_1D_s_twiddle}. As an example, we plot the spectrum \eqref{asymptotics_spec_twiddle} for~$\alpha = 2$ in Fig. \ref{fig:spec_cartoon_a2}.

In summary, we have constructed a phenomenological theory according to which $\delta C_2$ undergoes a phase-space cascade in both position and velocity space. At large $k$ ($\xi \gg 1$), the spectrum is phase-mixing-dominated and has a power-law scaling in $k$. At large $s$ ($\xi \ll 1$), the spectrum is turbulent-diffusion-dominated and has a power-law scaling in $s$. These two regimes are separated in phase space by the critical-balance region $\xi \sim 1$, where the linear and nonlinear time scales are comparable. The 1-D $k$ and $s$ spectra are dominated by contributions from this critical-balance region.

Since the collision operator is diffusive in velocity space and $\nu$ is assumed to be small, collisional dissipation must necessarily occur at fine velocity-space scales (large $s$). It is therefore unsurprising that a kinetic system with injection of a quadratic invariant exhibits a constant-flux cascade of that invariant in velocity space. However, there is no \textit{a priori} scale in position space where the dissipation must happen, so it is nontrivial that there also exists a cascade in position space. This inertial range in position space emerges due to the nonlinear field-particle interactions in \eqref{spec_eq_FL_noflux}, which mix position and velocity space~\cite{adkins2018solvable}. 

A reader interested in how the above results are obtained more rigorously should read Sections \ref{forced} and \ref{spec_self_similar}, where we solve \eqref{spec_eq_FL_noflux} properly for the spectrum and fluxes of $\delta C_2$. A reader interested only in the big picture can skip straight to Section \ref{discussion}.

\section{Phase-space fluxes} \label{forced}

We now verify the phenomenological theory presented in Section \ref{twiddle_theory} by solving \eqref{spec_eq_FL_noflux}. It is useful to write this equation in flux-gradient form:
\begin{equation} \label{spec_eq_FL}
    \frac{\partial \hat{F}}{\partial t} + \nabla \cdot \mathbf{\hat{\Gamma}} = \hat{S} - 2 \nu s^2 \hat{F},
\end{equation}
where $\nabla = \left(\partial/\partial k, \partial/\partial s \right)$ is a gradient operator in the~$(k,s)$ space, and the flux $\mathbf{\hat{\Gamma}} = \left(\hat{\Gamma}^k, \hat{\Gamma}^s \right)$ has components
\begin{equation} \label{gamma_s}
    \hat{\Gamma}^s = k \hat{F},
\end{equation}
which is clear from \eqref{spec_eq_FL_noflux}, and
\begin{equation} \label{gamma_k}
    \hat{\Gamma}^k = i \, \kappa s^2 \frac{1}{L} \int^{+\infty}_{-\infty} dr \, e^{-ikr} \frac{|r|^{\alpha}}{r} F(r,s).
\end{equation}
The latter expression can be derived using \eqref{Fourier_Transform_fracLap}, viz., by writing the Fourier transform of the nonlinear term as
\begin{equation} \label{Gamma_k_deriv}
    \kappa s^2 \, |r|^{\alpha}F(r,s) = -i r \left[ i \kappa s^2 \frac{|r|^{\alpha}}{r} F(r,s) \right].
\end{equation}
Inverse-Fourier transforming the term in brackets in \eqref{Gamma_k_deriv} yields \eqref{gamma_k} \cite{pozrikidis2018fractional}. In steady state, \eqref{spec_eq_FL} reads
\begin{equation} \label{final_eq_flux_ss}
     \nabla \cdot \mathbf{\hat{\Gamma}} = k\frac{\partial \hat{F}}{\partial s} + \kappa \, s^2 (-\Delta_k)^{\alpha/2}\hat{F} = \hat{S} - 2 \nu s^2 \hat{F}.
\end{equation}

In Sections \ref{const_flux} and \ref{const_flux_s}, we compute $\hat{\Gamma}^k$ and $\hat{\Gamma}^s$ integrated over $s$ and $k$, respectively. Then, in Section \ref{spec_self_similar}, we compute the spectrum $\hat{F}(k,s)$ and analyze the full flux $\mathbf{\hat{\Gamma}}$ in $(k,s)$ space. This order of presentation may seem awkward, but is, in fact, necessary. This is because the integrated fluxes inform us of the nature of the solution of \eqref{final_eq_flux_ss}. This solution is in turn needed to compute the 2-D flux in $(k,s)$ space.

\subsection{Constant flux in \texorpdfstring{$k$}{k}} \label{const_flux}
Let us integrate \eqref{final_eq_flux_ss} over all $s$. The $s$ flux vanishes at $s \rightarrow \pm \infty$, and we are left with an equation for $\hat{g}(k) = \int^{+\infty}_{-\infty} ds \, s^2 \hat{F}$, which satisfies
\begin{equation} \label{g_eq}
    \kappa (-\Delta_k)^{\alpha/2} \hat{g} = \mathcal{\hat{S}} - 2 \nu \hat{g},
\end{equation}
where $\mathcal{\hat{S}}(k) = \int^{+\infty}_{-\infty} ds \, \hat{S}(k,s)$. This integrated source $\mathcal{\hat{S}}$ injects $\hat{g}$, which is diffused by the nonlinear term and dissipated by collisions.

The source \eqref{source_sk} is peaked at low $k$, with characteristic width $L_E^{-1}$. To analyze the behavior of $\hat{g}(k)$ in the region $k L_E \gg 1$, we approximate $\mathcal{\hat{S}}(k) \approx (\varepsilon/L) \, \delta(k)$. Fourier transforming~\eqref{g_eq} and solving for $g(r)$, the Fourier transform of $\hat{g}(k)$, yields
\begin{equation} \label{g_r}
    g(r) = \frac{\varepsilon}{2 \pi \kappa}  \frac{1}{|r|^{\alpha} + \ell^{\alpha}_{\nu}},
\end{equation}
whence
\begin{equation} \label{g_k}
        \hat{g}(k) = \frac{1}{L} \int dr \, e^{-ikr} g(r) = \frac{\varepsilon}{2 \pi \kappa L} \int dr \, \frac{e^{-ikr}}{|r|^{\alpha} + \ell^{\alpha}_{\nu}}, 
\end{equation}
where we have defined the `Kolmogorov' (dissipation) scale as
\begin{equation} \label{kolmogorov_scale}
    \ell_{\nu} \equiv \left(\frac{2 \nu}{\kappa}\right)^{1/\alpha}.
\end{equation}
For $\alpha = 2$, \eqref{g_k} simplifies to
\begin{equation} \label{g_k_alpha2}
    \hat{g}(k) = \frac{\varepsilon}{2 \pi \kappa L} \int dr \, \frac{e^{-ikr}}{r^2 + \ell^2_{\nu}} =  \frac{\varepsilon \ell_{\nu} }{4 \nu L}e^{-|k| \ell_{\nu}}.
\end{equation}
For $\alpha < 2$, \eqref{g_k} does not have a simple closed-form expression, but it can be manipulated into an integral where the exponential in the integrand is decaying rather than oscillatory:
\begin{equation} \label{gk_branch}
    \hat{g}(k) = \frac{\varepsilon \ell_{\nu} }{2 \pi \nu L} \int^{\infty}_{0} dz \, \frac{\sin(\pi \alpha /2) \, z^{\alpha}}{z^{2\alpha} + 2z^{\alpha}\cos(\pi \alpha /2)+1} e^{-|k| \ell_{\nu}z}.
\end{equation}
This will be useful in what follows. Deriving \eqref{gk_branch} from~\eqref{g_k} requires some work, which is done in Appendix~\ref{branch_cut_calc_section}.

The solution \eqref{g_k} implies that the rate of $\delta C_2$ injection by the source equals the rate of $\delta C_2$ dissipation by collisions. Indeed, using \eqref{C2_Fhat_balance} and \eqref{g_k}, and converting the sums over $k$ into integrals, the collisional dissipation can be written in terms of $\hat{g}$:
\begin{align} \label{dissp_anom}
   & 2 \nu L \int dk \, \hat{g}(k) = 2 \nu \int dk \, \frac{\varepsilon}{2 \pi \kappa} \int dr \, \frac{e^{-ikr}}{|r|^{\alpha} + \ell^{\alpha}_{\nu}} \nonumber \\& = \varepsilon \, \ell_{\nu}^{\alpha} \int dr \, \delta(r) \frac{1}{|r|^{\alpha} + \ell^{\alpha}_{\nu}} = \varepsilon = L \iint dk \, ds \, \hat{S},
\end{align}
as expected. Since \eqref{g_k} is a Green's function solution to \eqref{g_eq}, it is straightforward to show that this balance also holds for arbitrary $ \mathcal{\hat{S}}(k)$. Importantly, \eqref{dissp_anom} applies even in the limit $\nu \rightarrow 0^{+}$; emergence of such finite collisional dissipation in the collisionless limit is known as a dissipative anomaly \cite{eyink2006onsager}. This result, although perhaps obvious from the steady state of \eqref{C2_Fhat_balance}, demonstrates constructively that the steady-state solution to \eqref{spec_eq_FL} is well defined in the collisionless limit.  

Using the solution \eqref{g_r}, we can now compute the 1-D $k$ flux of $\delta C_2$, viz., from \eqref{gamma_k},
\begin{equation} \label{kflux_g}
     L \int ds \, \hat{\Gamma}^k = i \kappa \int_{-\infty}^{+\infty} dr \, e^{-ikr} \frac{|r|^{\alpha}}{r} g(r).
\end{equation}
For $\alpha = 2$, this flux reduces to
\begin{equation} \label{flux_k_alpha2}
   L \int ds \, \hat{\Gamma}^k = - L \, \kappa \frac{\partial \hat{g} }{\partial k} = \, \mathrm{sgn}(k) \, \frac{\varepsilon }{2 } \, e^{-|k| \ell_{\nu}},
\end{equation}
where we used the solution \eqref{g_k_alpha2}. For $\alpha < 2$,

\begin{align} \label{flux_k_alpha}
   &  L \int ds \, \hat{\Gamma}^k  \nonumber  \\ & = \mathrm{sgn}(k) \, \frac{\varepsilon }{ \pi} \int^{\infty}_{0} dz \, \frac{\sin(\pi \alpha /2) \, z^{\alpha-1}}{z^{2\alpha} + 2z^{\alpha}\cos(\pi \alpha /2) + 1}  e^{-|k| \ell_{\nu} z}.
\end{align}
The derivation of \eqref{flux_k_alpha} can be found in Appendix \ref{branch_cut_calc_section}. When $k \ell_{\nu} \ll 1$, both \eqref{flux_k_alpha2} and \eqref{flux_k_alpha} are constant and equal to $\mathrm{sgn}(k) \, \varepsilon/2$ (in this limit, the integral over $z$ in \eqref{flux_k_alpha} is $\pi/2$, which is also shown in Appendix \ref{branch_cut_calc_section}). This result, one of the most important of this paper, means there is a constant-flux cascade in position space, viz., there exists an inertial range, $1/L_E \ll k \ll 1/\ell_{\nu}$, unaffected directly by forcing or collisions, where $\delta C_2$ is transferred from larger to smaller scales.

At $k \ell_{\nu} \gtrsim 1$, collisions become relevant. The dissipation rate at scales $ \gtrsim 1/k$ is
\begin{equation} \label{dissp_function}
\mathcal{\hat{D}}(k) = 2 \nu L \int^{+k}_{-k} dk' \hat{g}(k').
\end{equation}
Substituting \eqref{g_k_alpha2} or \eqref{gk_branch} into \eqref{dissp_function} gives, for $\alpha = 2$,
\begin{equation} \label{dissp_k_alpha2}
    \mathcal{\hat{D}}(k) = \frac{\varepsilon \ell_{\nu}}{2} \int^{+k}_{-k} dk' \, e^{-|k'| \ell_{\nu}} = \varepsilon(1-e^{-k \ell_{\nu}})
\end{equation}
or, for $\alpha < 2$,
\begin{equation} \label{dissp_k_alpha}
    \mathcal{\hat{D}}(k) = \frac{2 \varepsilon }{ \pi} \int^{\infty}_{0} dz \, \frac{\sin(\pi \alpha /2) \, z^{\alpha-1}}{z^{2\alpha} + 2z^{\alpha}\cos(\pi \alpha /2)+1} (1-e^{-k \ell_{\nu}z}).
\end{equation}
From \eqref{dissp_k_alpha2} and \eqref{dissp_k_alpha}, we see that the collisional dissipation is only order-unity when $k \ell_{\nu} \gtrsim 1$, i.e., below the Kolmogorov scale \eqref{kolmogorov_scale}, which, therefore, deserves the name that we have given it. Past $k \ell_{\nu} \gtrsim 1$, the dissipation balances the $\delta C_2$ injected by the forcing ($\mathcal{\hat{D}}(k) \simeq \varepsilon$ when $k\ell_{\nu} \gg 1$; derived in Appendix \ref{branch_cut_calc_section}). As discussed at the end of Section \ref{twiddle_theory}, because the collision operator is diffusive in velocity space, there is no \textit{a priori} scale in position space where the dissipation must happen. Rather, this dissipation range in position space forms because of the collisionless dynamics. Note that $\ell_{\nu} \rightarrow 0$ as $\nu \rightarrow 0$, so arbitrarily fine-scale structure in position space can be generated in the collisionless limit. Because of the constant-flux cascade, all of the $\delta C_2$ injected at large scales reaches the dissipation range, no matter how small $\ell_{\nu}$ is. 

\subsection{Constant flux in \texorpdfstring{$s$}{s}} \label{const_flux_s}

Let us now integrate \eqref{final_eq_flux_ss} over all $k$. The $k$ flux vanishes at $k \rightarrow \pm \infty$, resulting in the following equation for the $s$ flux:

\begin{equation} \label{sflux_eq}
    \frac{\partial}{\partial s} \int dk \, \hat{\Gamma}^s = \int dk \, \hat{S} - 2 \nu s^2 \int dk \, \hat{F}.
\end{equation}
Unfortunately, unlike \eqref{g_eq}, this equation is not closed and cannot be explicitly solved without knowing the spectrum itself. However, we can still learn a key lesson from it. In view of \eqref{source_sk} and \eqref{source_sk_Maxwellian}, the characteristic width over which $\int dk \, \hat{S}$ falls off in $s$ is $1/v_{\mathrm{th}}$. But for small $\nu$, collisions can only be relevant when $s$ is large. Balancing the phase-mixing term with the collision term in \eqref{final_eq_flux_ss} tells us that collisions will start to matter when
\begin{equation} \label{bal_fs_coll}
    \frac{k}{s} \sim \nu s^2.
\end{equation}
We know from Section \ref{const_flux} that collisional dissipation will start to occur around $k\ell_{\nu} \sim 1$, so taking $k \sim 1/\ell_{\nu}$ in~\eqref{bal_fs_coll} gives us the collisional velocity scale
\begin{equation} \label{u_nu_k_nu}
    u_{\nu} \sim \left( \frac{\nu^{\alpha+1}}{\kappa}\right)^{1/3 \alpha}.
\end{equation}

When $s u_{\nu} \ll 1$, we expect that we can drop the collision term in \eqref{sflux_eq}. Integrating the remaining terms in the equation from $-s$ to $s$, where $1/v_{\mathrm{th}} \ll s \ll 1/u_{\nu}$, yields, using \eqref{const_inj},

\begin{align} \label{s_flux_integration}
    & \int dk \, \hat{\Gamma}^s(s) - \int dk \, \hat{\Gamma}^s(-s) \nonumber \\ &=  \int^{+s}_{-s} ds' \, \int^{+\infty}_{-\infty} dk \, \hat{S}(k,s') \approx \mathrm{sgn}(s) \frac{\varepsilon}{L}.
\end{align}
Because the distribution function is real, the Fourier spectrum satisfies
\begin{equation} \label{reality_conditions}
    \hat{F}(k, -s) = \hat{F}(-k, s).
\end{equation}
Together with the definition \eqref{gamma_s} of $\hat{\Gamma}^s$, this implies that $\int dk \, \hat{\Gamma}^s(-s) = -\int dk \, \hat{\Gamma}^s(s)$, and so \eqref{s_flux_integration} becomes
\begin{equation} \label{s_flux_constant}
    L \int dk \, \hat{\Gamma}^s(s) = \mathrm{sgn}(s) \, \frac{\varepsilon}{2}
\end{equation}
for $1/v_{\mathrm{th}} \ll s \ll 1/u_{\nu}$. There is, therefore, also an inertial range in $s$ where there is a constant flux of $\delta C_2$ from small to large $|s|$. Because the collision operator can only be relevant at large $|s|$ when $\nu$ is small, the existence of a forced steady state does indeed require such a constant-flux inertial range in $s$.

We note that this result holds even in the absence of nonlinearity. Linear phase mixing can lead to the development of arbitrarily fine scales in velocity space, albeit at the price of a diverging $\delta C_2$ as $\nu \rightarrow 0^{+}$ \cite{kanekar2015fluctuation}. However, nonlinearity is required for there to be a cascade in position space: note from \eqref{kolmogorov_scale} that $\ell_{\nu} \rightarrow \infty$ as $\kappa \rightarrow 0$. Indeed, the solution of \eqref{g_eq} for $\kappa = 0$ is simply $\hat{g} = \mathcal{\hat{S}}(k)/2 \nu$. The Vlasov equation is linear in this limit, so the spatial Fourier components of the distribution function are uncoupled from one another, and the $k$ dependence of the Fourier spectrum is then simply set by the source.

\section{Phase-space spectrum} \label{spec_self_similar}

We now compute the spectrum in \eqref{final_eq_flux_ss}. From Sections \ref{const_flux} and \ref{const_flux_s}, we know that our model exhibits a phase-space entropy cascade in which the rate of $\delta C_2$ injection by the forcing at large scales is balanced by collisional dissipation at small scales, with the distribution function developing arbitrarily small scales in phase space in the limit $\nu \rightarrow 0^{+}$. This results in an inertial range in $k$ ($s$) where the flux of $\delta C_2$ integrated over velocity (position) wavenumbers is constant. Therefore, in this inertial range, viz., for $1/L_E \ll k \ll 1/\ell_{\nu}$ and $1/v_{\mathrm{th}} \ll s \ll 1/u_{\nu}$, we can meaningfully consider \eqref{final_eq_flux_ss} in the absence of the forcing and collisional terms, viz.,
\begin{equation} \label{final_eq_inertial_range}
    k\frac{\partial \hat{F}}{\partial s} = -\kappa s^2 (-\Delta_k)^{\alpha/2}\hat{F},
\end{equation}
and seek a solution for the spectrum that supports constant fluxes in $k$ and $s$.

\subsection{Self-similar inertial-range solution} \label{self-similar-solution}

Because of the reality condition \eqref{reality_conditions}, we only need to solve \eqref{final_eq_inertial_range} for $\hat{F}(k,s)$ on half of the $(k,s)$ plane. We choose to solve for the spectrum in the upper half-plane, $-\infty < k < \infty$ and $s \geq 0$ \footnote{This solution can be directly compared to earlier results obtained in terms of the Hermite numbers $m$, as the Hermite transform is analogous to a Fourier transform when $m \gg 1$, with $m \sim s^2/2$ \cite{adkins2018solvable}.}.

To deal with the fractional Laplacian, we Fourier transform this equation in $k$ space. Using \eqref{Fourier_Transform_fracLap}, we find that $F(r,s)$ satisfies
\begin{equation} \label{Frs_fract}
     \frac{\partial}{\partial r} \frac{\partial}{\partial s}F = -i \kappa s^2 |r|^{\alpha} F.
\end{equation}
Because $F(r,s)$ is the Fourier transform of $\hat{F}(k,s)$, which is purely real, $F(r,s)$ must satisfy the reality conditions $F(-r,s) = F^*(r,s)$. We therefore only have to solve \eqref{Frs_fract} for $r > 0$, for which $|r| = r$.

The inertial-range solution to \eqref{Frs_fract} that does not depend on details of the forcing or dissipation ranges is a similarity solution \cite{barenblatt1996scaling}:
\begin{equation} \label{SimSol}
    F = \varepsilon \, \kappa^{-1/(\alpha + 1)} s^{-\beta}\phi(y), \quad y = (\kappa s^3)^{1/(\alpha+1)} \, r,
\end{equation}
where $\beta$ can be constrained by the fact that the flux $L \int dk \, \hat{\Gamma}^s$ must be constant in the inertial range, as per~\eqref{s_flux_constant}:
\begin{align} \label{flux_constraint}
 & L \int dk \, \hat{\Gamma}^s \nonumber \\  &= \varepsilon \, s^{-\beta + 3/(\alpha + 1)} \int_{-\infty}^{+\infty} & d \xi \, \xi \, \int_{-\infty}^{+\infty} dy \, e^{-i\xi y} \phi(y) = \frac{\varepsilon}{2}, \nonumber \\ & \implies \beta = \frac{3}{\alpha + 1}.
\end{align}
The integration variable dual to $y$,
\begin{equation} \label{xi}
    \xi = \frac{k}{(\kappa \, s^3)^{1/(\alpha+1)}},
\end{equation}
has previously appeared in \eqref{ratio_time_scale} as the ratio of the turbulent-diffusion and phase-mixing time scales.

Substituting \eqref{SimSol} into \eqref{Frs_fract} gives us an ordinary differential equation for $\phi$:
\begin{equation} \label{phi_eq}
    \frac{d^2 \phi}{dy^2} + i\frac{(\alpha+1)}{3}y^{\alpha-1}\phi = 0.
\end{equation}
To solve this equation, consider the transformation
\begin{equation}
    \phi = \sqrt{y} \, g(z), \quad z = \frac{2}{\sqrt{3(\alpha+1)}}e^{-i \pi /4}y^{(\alpha + 1)/2}.
\end{equation}
Then, \eqref{phi_eq} becomes a modified Bessel equation \cite{abramowitz1964handbook}:
\begin{equation}
    z^2 \frac{d^2 g}{dz^2} + z \frac{d g}{dz} -\left[z^2 + \frac{1}{(\alpha+1)^2}\right]g = 0.
\end{equation}
Therefore, the solution to \eqref{phi_eq} that vanishes at $y \rightarrow \infty$~is
\begin{align} \label{phi}
    \phi(y) =  \Lambda \, \sqrt{y} \, K_{1/(\alpha + 1)}\left(\frac{2}{\sqrt{3(\alpha+1)}}e^{-i \pi /4}y^{(\alpha+1)/2}\right),
\end{align}
where $y > 0$, $K$ is the modified Bessel function of the second kind, and $\Lambda$ is a constant. The reality condition $\phi(y) = \phi^*(-y)$ constrains us to pick the phase of $\Lambda$ so that the real part of $\phi$ is even in $y$ and the imaginary part is odd. This is accomplished by setting
\begin{equation}
    \Lambda = \sigma (e^{-i \pi /4})^{1/(\alpha+1)},
\end{equation}
where $\sigma > 0$ is real and determined by the constraint~\eqref{flux_constraint}. For generic $\alpha$, we are unable to evaluate the integrals in \eqref{flux_constraint} analytically, to compute $\sigma$, although it is straightforward to see that $\sigma$ is finite. For $\alpha = 1$,~$\sigma$ is easily computed: this calculation can be found in Appendix \ref{alpha1_calc_section}.

Using the solution \eqref{phi} for $\phi$, we inverse-Fourier transform back to $k$ space to obtain,
\begin{align} \label{Fks}
 &\hat{F}(k,s) = \nonumber \\ & 2 \, \varepsilon \, L^{-1} \kappa^{-2/(\alpha + 1)} s^{-6/(\alpha + 1)} \operatorname{Re} \int_{0}^{\infty} dy \, e^{-i\xi y} \phi(y), 
\end{align}
where $\xi$ is given by \eqref{xi}, and we have used the reality condition for $\phi$. As we stated at the beginning of this calculation, this expression is valid only for $s \geq 0$; the spectrum for negative $s$ can be found by combining \eqref{reality_conditions} and \eqref{Fks}. For generic $\alpha$, we are unable to simplify \eqref{Fks} further. For $\alpha = 1$, \eqref{Fks} has a simple closed form, derived in Appendix \ref{alpha1_calc_section}. For $\alpha = 2$, the spectrum can be written in terms of incomplete Gamma functions: see \cite{adkins2018solvable}. In our terms, \eqref{phi} for $\alpha = 2$ reduces to
\begin{equation}
    \phi(y) = \sigma \pi \sqrt{3} \mathrm{Ai} \left( e^{-i \pi/6} y\right),
\end{equation}
where $\mathrm{Ai}$ is the Airy function. This is clear by the fact that for $\alpha = 2$, \eqref{phi_eq} is an Airy equation.

\subsection{2-D spectrum} \label{PD_section}

\begin{figure}
	\centering
	\includegraphics[width=1\linewidth]{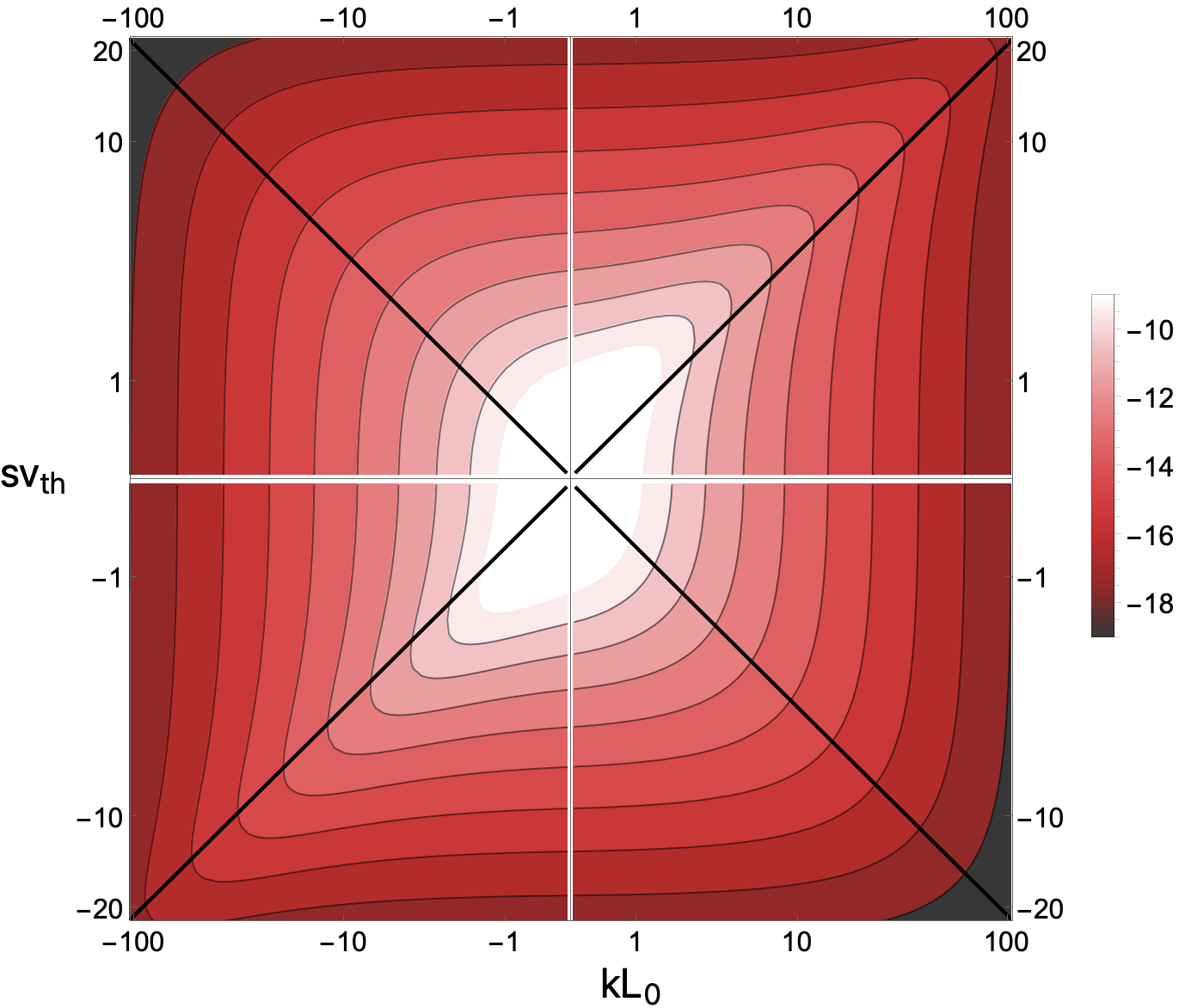}
	\caption{\label{fig:spec_a1} Contour plot of $\log L_0^2 \, v_{\mathrm{th}}^{6/(\alpha + 1)} \hat{F}(k L_0 , s v_{\mathrm{th}})$ vs. $(k L_0 ,s v_{\mathrm{th}})$, for $\alpha = 1$. Note that the spectrum for $s < 0$ is given by the reality condition \eqref{reality_conditions}, $\hat{F}(k,s) = \hat{F}(-k,-s)$. The black lines are the critical-balance lines~\eqref{CB}, $s v_{\mathrm{th}} = \pm |k L_0|^{2/3}$. In order to use logarithmic axes, we do not plot the spectrum along the $k = 0$ and $s = 0$ axes. We have also not plotted the spectrum near the $(k,s)$ origin, as the similarity solution \eqref{Fks} diverges there.}
\end{figure}

To show what \eqref{Fks} looks like, we present contour plots of the normalized spectrum $L_0^2 \, v_{\mathrm{th}}^{6/(\alpha + 1)} \hat{F}(k L_0 , s v_{\mathrm{th}})$ for $\alpha = 1$ (for which an explicit expression can be found in Appendix \ref{alpha1_calc_section}) in Fig.~\ref{fig:spec_a1}. We use normalized units $s v_{\mathrm{th}}$ and $k L_0$, where
\begin{equation} \label{L0}
    L_0 = \left( \frac{v_{\mathrm{th}}^3}{\kappa} \right)^{1/(\alpha+1)}.
\end{equation}
This length scale is a natural choice, because the similarity variable \eqref{xi} in these units is
\begin{equation} \label{xi_norm}
    \xi = \frac{k}{(\kappa \, s^3)^{1/(\alpha+1)}} = \frac{k L_0}{\left(s v_{\mathrm{th}} \right)^{3/(\alpha+1)}},
\end{equation}
and, therefore, the spectrum, up to its amplitude, is independent of $\kappa$.

For $\xi \sim 1$, the spectrum is a nontrivial function of $k$ and $s$. For $\xi$ small or large, it has asymptotics given by~\eqref{asymptotics_spec_twiddle} (computed below), confirming the phenomenological theory presented in Section \ref{twiddle_theory}. As discussed in Section \ref{twiddle_theory}, the distinction between $\xi$ small and large can be understood in terms of competition between the phase mixing and turbulent diffusion for control of the phase-space cascade.

To compute the asymptotics \eqref{asymptotics_spec_twiddle} from the full solution~\eqref{Fks}, the $\xi \ll 1$ limit can be found by simply setting $\xi = 0$ in~\eqref{Fks}. For the $\xi \gg 1$ limit, we use the following result from Fourier analysis. Suppose $u(y)$ is a function that is smooth for $y > 0$ and has the scaling $u(y) \sim y^{\lambda-1}$ as ${y \rightarrow 0^{+}}$, where $\lambda > 0$. Then
\begin{equation} \label{fourier_result}
    \int_{0}^{\infty} dy \, e^{-i\xi y} u(y) = \frac{\Gamma(\lambda)}{(i \xi)^{\lambda}} + o\left( \xi^{-\lambda} \right)
\end{equation}
as $\xi \rightarrow \infty$ \cite{saichev2018distributions}. Note that \eqref{phi} has the series
\begin{equation} \label{phi_series}
    \phi(y) = A - B (e^{-i \pi/2})^{1/(\alpha + 1)} y + \mathcal{O}(y^{\alpha + 1})
\end{equation}
as $y \rightarrow 0^{+}$, where $A$ and $B$ are real and positive constants. Combining \eqref{fourier_result} and \eqref{phi_series}, one finds that the contribution to the integral in \eqref{Fks} from the constant part of \eqref{phi_series} is purely imaginary and so vanishes. The next-order term from the linear part of \eqref{phi_series} has a real component; the integral is therefore $\mathcal{O}(\xi^{-2})$ (since $\lambda = 2$), hence $\hat{F}(k,s) \propto k^{-2}$ as $\xi \rightarrow \infty$.

Note that, the 1-D $k$ and $s$ spectra, which can be found by integrating out the similarity variable $\xi$ in \eqref{Fks}, agree with \eqref{spec_1D_k_twiddle} and \eqref{spec_1D_s_twiddle}.

Before continuing, it is instructive to assess the region of validity of \eqref{Fks} in the $(k,s)$ plane. This spectrum is infrared-divergent and thus breaks down as $(k,s) \rightarrow (0,0)$; information about the functional form of the source \eqref{source_sk}, which would regularize the spectrum at long wavelengths, is lost by construction in the similarity solution (however, \eqref{Fks} does contain information about the flux from the source via the constraint \eqref{flux_constraint}). Of course, the inertial-range spectrum is no longer valid in the region where the source \eqref{source_sk} is concentrated, viz., when $sv_{\mathrm{th}} \lesssim 1$ and $k L_{E} \lesssim 1$. For these reasons, we have not plotted the spectrum near the origin in the $(k,s)$ plane in Fig.~\ref{fig:spec_a1}, and likewise for the fluxes plotted in the following sections \footnote{The fact that the inertial-range spectrum \eqref{Fks} is scale-free and has no infrared cutoff means that $v_{\mathrm{th}}$ in \eqref{L0} and \eqref{xi_norm} is formally an arbitrary velocity normalization for \eqref{Fks}. However, to emphasize that the true spectrum has an outer scale at $s v_{\mathrm{th}} \lesssim 1$, we still normalize $s$ by $v_{\mathrm{th}}$ in \eqref{Fks}.}.

In addition to the solution \eqref{Fks} lacking an outer scale, the approximation that the nonlinear term is a fractional Laplacian to lowest order also breaks down when~$k L_{E} \lesssim~1$. This is clear in the derivation of the nonlinear term in Appendix \ref{app_frac_Lap_deriv}, where the fractional Laplacian emerges as the lowest-order term when $k L_{E} \gg 1$. Yet, away from $s = 0$, \eqref{Fks} is, in fact, continuous across $k = 0$. By dropping the finite-$ k L_{E}$ corrections in \eqref{spec_eq_FL_noflux}, we have shrunk the boundary layer $k L_{E} \lesssim 1$ to the point $k = 0$~\footnote{For the source term to be non-zero, it is important that we do not take the limit $L_{E} \rightarrow \infty$, otherwise $\varepsilon \rightarrow 0$. This is because $\varepsilon \propto D_0^{-1/2} \propto L^{-\alpha/2}_{E}$, as can be seen in \eqref{source_C2_time_generic}. Physically, turbulent diffusion is a large-scale effect and so is dominated by the largest electric-field scales. Note that any meaningful analysis of the spectrum in the case of $L_{E} \rightarrow \infty$ would require a time-dependent solution to \eqref{spec_eq_FL_noflux} \cite{eyink2000self,Chaves2001,Chertkov2003}}.

We can also now address the concern of the lack of locality in $k$ space of the source term \eqref{source_sk} and whether dropping this term in the inertial range was justified, as discussed at the end of Section \ref{Fourier_eq}. Consider the two asymptotic regions, $\xi \ll 1$ and $\xi \gg 1$ (assuming also $sv_{\mathrm{th}} \gg 1$, $s u_{\nu} \ll 1$, and $k \ell_{\nu} \ll 1$). In the former region, the nonlinear term is dominant over the phase-mixing term, as per \eqref{homog_frac}. Balancing the nonlinear term with~\eqref{source_sk} yields an inhomogeneous contribution to $\hat{F} \propto \langle \hat{f} \rangle^2$. Since $sv_{\mathrm{th}} \gg 1$, this term is strongly suppressed, e.g., exponentially so in the case of a Maxwellian initial condition [see~\eqref{source_sk_Maxwellian}], so this term is negligible compared to the $\xi \ll 1 $ asymptotic in \eqref{asymptotics_spec_twiddle}. In the the latter region ($\xi \gg 1$), the phase-mixing term is dominant over the nonlinear term, as per \eqref{linear_constant_flux}. Balancing the phase-mixing term with \eqref{source_sk} yields an inhomogeneous contribution to $\hat{F} \propto \hat{D}(k)/k$, which is $\propto k^{-(2+\alpha)}$ when $kL_E \gg 1$. This contribution is subdominant to the homogeneous part of the spectrum, which scales like $k^{-2}$ when $\xi \gg 1$, viz., \eqref{asymptotics_spec_twiddle}. We cannot explicitly estimate the inhomogeneous contribution to the spectrum from the source in the $\xi \sim 1$ region, but the above analysis suggests it should be subdominant to~\eqref{Fks}. Therefore, even though the source is multiscale in~$k$, there is still an inertial range in the position space.

Finally, we observe that the inertial-range solution \eqref{Fks} extends to infinity, consistent with $k_{\nu} \sim 1/\ell_{\nu} \rightarrow \infty$ and  $s_{\nu} \sim 1/u_{\nu} \rightarrow \infty$ as $\nu \rightarrow 0^{+}$, viz., \eqref{kolmogorov_scale} and \eqref{u_nu_k_nu}. A finite $\nu$ will introduce finite collisional scales $k_{\nu}$ and $s_{\nu}$, such that collisions cut off the spectrum when $k \gtrsim k_{\nu}$ and $s \gtrsim s_{\nu}$.

\subsection{2-D flux: phase-space circulations} \label{Flux_analysis}

To gain insight about the pathways in phase space taken by $\hat{F}$ from injection to dissipation scales, it is informative to examine the vector flux $\mathbf{\hat{\Gamma}}$, which, in terms of the similarity solution \eqref{Fks}, has the components~\eqref{gamma_k},
\begin{equation} \label{gamma_k_phi}
    \hat{\Gamma}^{k}(k,s) = -2 \, \varepsilon \, L^{-1} s^{-1} \operatorname{Im} \int^{\infty}_{0} dy \, e^{-i \xi y} y^{\alpha-1} \phi(y),
\end{equation}
and \eqref{gamma_s},
\begin{align} \label{gamma_s_phi}
    &\hat{\Gamma}^{s}(k,s) \nonumber \\  & = 2 \, \varepsilon \, L^{-1} \kappa^{-1/(\alpha + 1)} s^{-3/(\alpha + 1)} \xi \operatorname{Re} \int^{\infty}_{0} dy \, e^{-i \xi y} \phi(y).
\end{align}
To obtain \eqref{gamma_k_phi}, we used \eqref{SimSol} and changed variables from $r$ to $y$ in the integral. Note that these expressions are valid only for $s \geq 0$. For $s < 0$, combining \eqref{gamma_s}, \eqref{gamma_k}, and \eqref{reality_conditions}, we have that
\begin{equation} \label{flux_negative_s}
\mathbf{\hat{\Gamma}}(k, -s) = -\mathbf{\hat{\Gamma}}(-k,s).
\end{equation}

As was the case for the Fourier spectrum, the flux is a nontrivial function of $k$ and $s$ for $\xi \sim 1$, but can be simplified when $\xi$ is small or large. The asymptotics of the $k$ and $s$ components of the flux, which we derive below, are
\begin{align} \label{asymptotics_k_flux}
    &\hat{\Gamma}^{k}(k,s) \sim \varepsilon \, L^{-1} \nonumber \\  &\times
\begin{cases} 
      -s^{-1}, \quad & \xi \ll 1, \\
      \textrm{sgn}(k) \, \kappa^{\alpha/(\alpha + 1)} |k|^{-\alpha} s^{(2\alpha -1)/(\alpha + 1)}, \quad & \xi \gg 1, \quad \alpha < 2, \\
      \kappa \, k^{-3} s^2, \quad & \xi \gg 1, \quad \alpha = 2.
   \end{cases}
   \end{align}
and
\begin{equation} \label{asymptotics_s_flux}
    \hat{\Gamma}^{s}(k,s) \sim \varepsilon \, L^{-1} \,
\begin{cases} 
      \kappa^{-2/(\alpha + 1)} \, k \, s^{-6/(\alpha + 1)}, \quad & \xi \ll 1, \\
       k^{-1}, \quad & \xi \gg 1,
   \end{cases}
\end{equation}
respectively. Here, we have retained the signs of terms as well as dimensional factors in \eqref{asymptotics_k_flux} and \eqref{asymptotics_s_flux}, but not order-unity constants. To evaluate the asymptotics of the fluxes for~$s < 0$, these expressions must be combined with \eqref{flux_negative_s}.

We can derive these results in the same way as we did the asymptotics of the spectrum in Section \ref{PD_section}. The asymptotics \eqref{asymptotics_s_flux} for $\hat{\Gamma}^{s}$ come directly from combining~\eqref{gamma_s} and \eqref{asymptotics_spec_twiddle}. For $\hat{\Gamma}^{k}$, the $\xi \ll 1$ expansion in~\eqref{asymptotics_k_flux} comes from evaluating \eqref{gamma_k_phi} at $\xi = 0$ (note that the integral is positive). For $\xi \gg 1$, we can combine \eqref{fourier_result}, \eqref{phi_series}, and \eqref{gamma_k_phi}. This gives, to lowest order, as $\xi \rightarrow \infty$,
\begin{equation} \label{asym_k_flux_alpha_not_2}
    \hat{\Gamma}^{k} \simeq 2 \, \varepsilon \, L^{-1} s^{-1} A \, \Gamma(\alpha) \sin\left( \frac{\pi \alpha}{2} \right) \textrm{sgn}(k) \, |\xi|^{-\alpha}.
\end{equation}
The lowest-order $k$ flux vanishes when $\alpha = 2$, so \eqref{asym_k_flux_alpha_not_2} only gives the $\xi \gg 1$ limit in \eqref{asymptotics_k_flux} for $\alpha < 2$. We need to go to next order for $\alpha = 2$, which yields
\begin{equation}
    \hat{\Gamma}^{k} \simeq 2 \, \varepsilon \, L^{-1} s^{-1} B \, \Gamma(3) \sin\left( \frac{\pi}{3} \right) \textrm{sgn}(k) \, |\xi|^{-3},
\end{equation}
giving the $\xi \gg 1$, $\alpha = 2$ asymptotic in \eqref{asymptotics_k_flux}.

\begin{figure}
	\centering
	\includegraphics[width=1\linewidth]{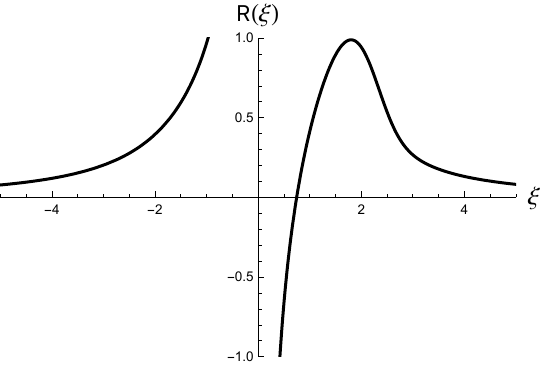}
	\caption{\label{fig:a1_R_xi} The ratio \eqref{flux_alpha2_ratio} versus $\xi$ for $\alpha = 2$, calculated by numerically integrating \eqref{Fks}. The point where $R$ vanishes, $\xi = \xi_{2} \simeq 0.747$, corresponds to $\hat{\Gamma}^{k} = 0$.}
\end{figure}

\begin{figure*}
	\centering
	\includegraphics[width=1\linewidth]{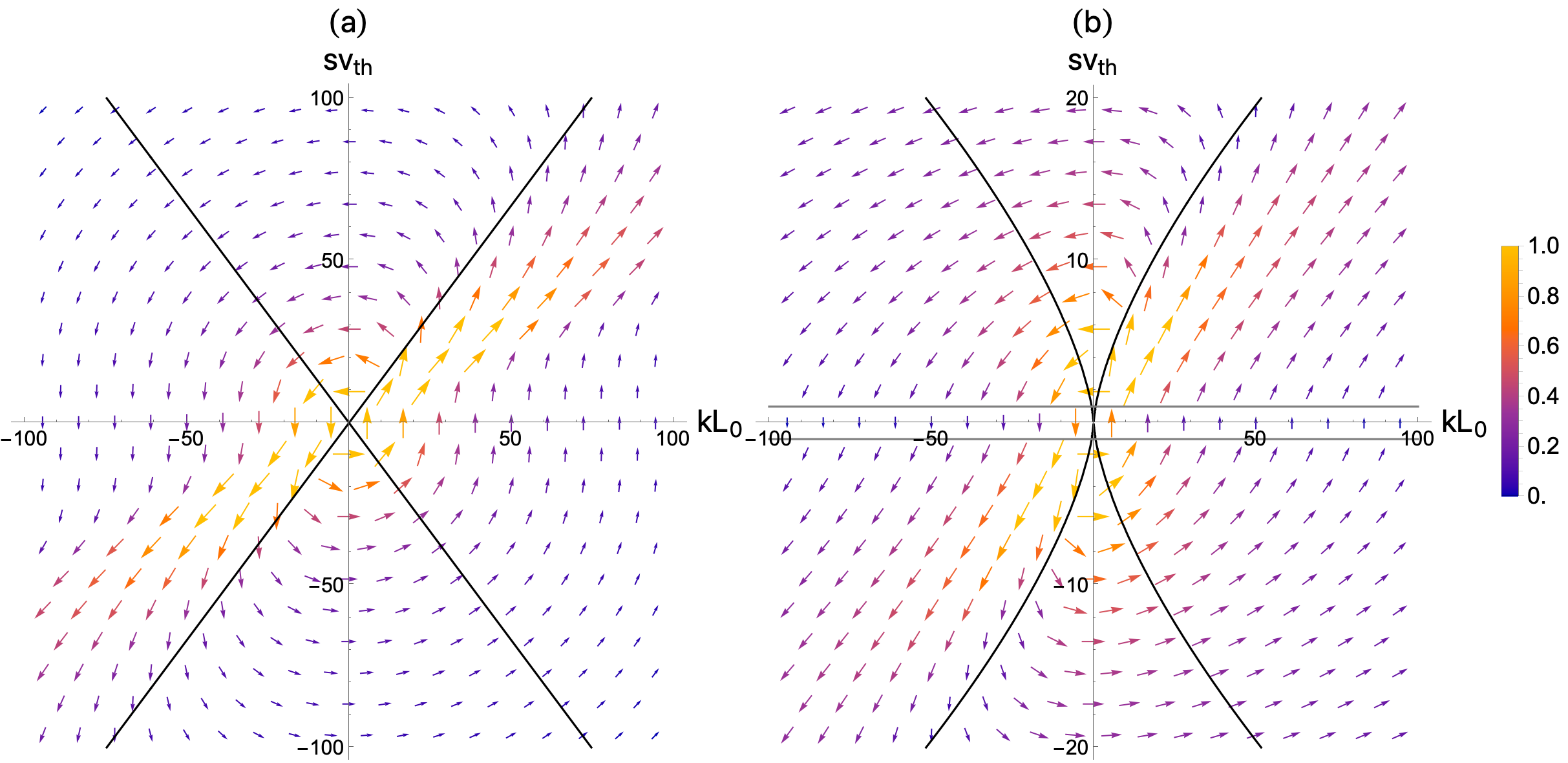}
	\caption{\label{fig:a1_flux_ks_alpha2_alpha1} (a) 2-D nondimensionalized flux for $\alpha = 2$, with $k$ component $L_0^4 \hat{\Gamma}^{k}(k L_0,s v_{\mathrm{th}})$ and $s$ component $L_0^2 v_{\mathrm{th}}^{6/(\alpha + 1)} \hat{\Gamma}^{s}(k L_0,s v_{\mathrm{th}})$, calculated by integrating \eqref{Fks} numerically. For emphasis, we specify the magnitude of the flux by both the length and color of the vectors. The black lines are the critical-balance lines~\eqref{CB} for these parameters, $s v_{\mathrm{th}} = \pm k L_0/\xi_2 $, where the slope $\xi_2 \simeq 0.747$ is chosen so that $\hat{\Gamma}^{k}$ vanishes along the line in the top right and bottom left quadrants. (b) The same as (a) but for $\alpha = 1$. Analytical expressions for the components of the flux in this case can be found in Appendix~\ref{alpha1_calc_section}, viz.,~\eqref{k_flux_a1} and \eqref{s_flux_a1}. The black curves are the critical-balance lines~\eqref{CB} for these parameters, $s v_{\mathrm{th}} = \pm |k L_0/\xi_1|^{2/3} $, where $\xi_1 = 1/\sqrt{3}$ is chosen so that $\hat{\Gamma}^{k}$ vanishes along the curve in the top right and bottom left quadrants [as can be seen in \eqref{k_flux_a1}]. The gray lines are $s v_{\mathrm{th}} = \pm 1$, bounding the only region, $| s v_{\mathrm{th}} | \ll 1$, where the flux is phase-mixing-dominated (apart from the critical-balance curve).}
\end{figure*}

We first discuss the Batchelor case ($\alpha = 2$), where
\begin{equation} \label{fluxBatchelor}
    \left(\hat{\Gamma}^k, \hat{\Gamma}^s \right) = \left(-\kappa s^2 \frac{\partial \hat{F}}{\partial k}, k \hat{F} \right).
\end{equation}
To work out by what physical mechanism the phase-space cascade is enabled in different parts of the $(k,s)$ plane, it is useful to compare the ratio of the $k$ and $s$ fluxes. Using the normalizations from Section \ref{PD_section}, we have that the ratio of the dimensionless fluxes is
\begin{equation} \label{flux_alpha2_ratio}
 R = \frac{L_0^4 \hat{\Gamma}^{k}(k L_0,s v_{\mathrm{th}})}{L_0^2 v_{\mathrm{th}}^{2} \hat{\Gamma}^{s}(k L_0,s v_{\mathrm{th}})} = -\frac{1}{\xi}\frac{\operatorname{Im} \int^{\infty}_{0} dy \, e^{-i \xi y} y \phi(y)}{ \operatorname{Re} \int^{\infty}_{0} dy \, e^{-i \xi y} \phi(y) },
\end{equation}
which is a function solely of $\xi$. It has the asymptotics
\begin{equation} \label{asymptotics_flux_alpha2_ratio}
 R \sim
\begin{cases} 
        -\xi^{-1}, \qquad \qquad & \xi \ll 1, \\
       \xi^{-2}, \qquad \qquad & \xi \gg 1.
   \end{cases}
\end{equation}
Thus, when $\xi \ll 1$, the flux is diffusion-dominated (dominated by its $k$ component), and when $\xi \gg 1$, the flux is phase-mixing-dominated (dominated by its $s$ component). When $\xi \sim 1$, the two fluxes are comparable; this can be seen in the plot of $R$ in Fig.~\ref{fig:a1_R_xi}. The regions of the dominance of the two fluxes are, therefore, separated in phase space by the critical-balance line~\eqref{CB}, the same as the phase-mixing and turbulent-diffusion time scales, viz., \eqref{ratio_time_scale}.

We plot the (nondimensionalized) vector flux \eqref{fluxBatchelor} in Fig.~\ref{fig:a1_flux_ks_alpha2_alpha1}(a). Using the asymptotics \eqref{asymptotics_flux_alpha2_ratio} and noting the signs of the fluxes in \eqref{asymptotics_k_flux} and \eqref{asymptotics_s_flux}, we see that in the diffusion-dominated region ($\xi \ll 1$), $\hat{\Gamma}^{k}$ is negative when $s > 0$ and positive when $s < 0$, and in the phase-mixing-dominated region ($\xi \gg 1$), $\hat{\Gamma}^{s}$ is positive when~$k > 0$ and negative when~$k < 0$. The flux therefore gives rise to counterclockwise circulation of $\delta C_2$ in $(k,s)$ space. The sign changes in the components of the flux that enable this circulation occur at $\xi = \xi_2 \simeq 0.747$, where $R$ has a zero (as can be seen in Fig.~\ref{fig:a1_R_xi}), and at~$\xi = 0$, below (above) which $R$ diverges positively (negatively). The first point corresponds to $\hat{\Gamma}^{k}$ changing sign in the top right and bottom left quadrants, while the second point corresponds to $\hat{\Gamma}^{s}$ changing sign from positive to negative between the top right and top left quadrants, as well as between the bottom left and bottom right quadrants. This latter effect occurs because, when $\mathrm{sgn}(ks) = - 1$, perturbations are phase unmixed rather than phase mixed, being advected to low $|s|$ rather than high $|s|$ \cite{schekochihin2016phase,adkins2018solvable}. The phase-unmixing modes are a stochastic instantiation of the plasma-echo effect \cite{gould1967plasma,malmberg1968plasma}.

While the phase unmixing does undo the phase mixing, we will see in Section \ref{sec_SA_flux} that the flux of the phase-mixing modes outweighs that of the phase-unmixing ones, thus enabling the constant-flux cascade. In Fig.~\ref{fig:a1_flux_ks_alpha2_alpha1}(a), this manifests in the fact that the circulation swirls outward. Indeed, in the top right (bottom left) quadrant, below (above) the line $\xi = \xi_2$ where $\hat{\Gamma}^{k} = 0$, there is a flux of $\delta C_2$ to both high $|k|$ and high $|s|$ simultaneously, toward the dissipation wavenumbers $k_{\nu} \sim 1/\ell_{\nu}$ and $s_{\nu} \sim 1/u_{\nu}$. 

\subsection{Non-local transport} \label{Flux_analysis_non_local}

We now examine the $\alpha < 2$ cases. The important difference compared to the Batchelor regime is that the $k$ flux is now non-local in $k$ space. Note that $\hat{\Gamma}^{k}$ can be written as \cite{pozrikidis2018fractional} \footnote{When $\alpha = 1 $, \eqref{gamma_k_nonlocal} is proportional to a Hilbert transform \cite{king_2009} in $k$ space. In plasma kinetics, non-local fluxes given by Hilbert transforms also show up in the context of Landau-fluid closures; viz., in the Hammett-Perkins closure \cite{hammett1990fluid}, the heat flux is proportional to the Hilbert transform of the temperature (in position space).}
\begin{equation} \label{gamma_k_nonlocal}
    \hat{\Gamma}^{k} = \kappa s^2 \frac{c_{\alpha}}{\alpha} \int^{\infty}_{0} dp \frac{\hat{F}(k-p,s)-\hat{F}(k+p,s)}{p^{\alpha}},
\end{equation}
where $c_{\alpha}$ is given by \eqref{calpha}. The derivation of this expression is given in Appendix \ref{derivation_gamma_k_nonlocal}. The interpretation of~\eqref{gamma_k_nonlocal} is that `particles' (parcels of $\delta C_2$) cross the point~$(k,s)$ from points $(k \pm p,s)$, with cumulative probability~$\propto p^{-\alpha}$. The particles undergo Lévy flights in $k$ space, so the flux at a point $(k,s)$ receives contributions not just from nearby particles taking small jumps but also from faraway ones taking large jumps.

The net effect is an enhancement of $\hat{\Gamma}^{k}$ compared to~$\hat{\Gamma}^{s}$. For these cases, the ratio $R$ of the nondimensionalized fluxes is
\begin{align} \label{flux_alpha_ratio}
 R &= \frac{L_0^4 \hat{\Gamma}^{k}(k L_0,s v_{\mathrm{th}})}{L_0^2 v_{\mathrm{th}}^{6/(\alpha + 1)} \hat{\Gamma}^{s}(k L_0,s v_{\mathrm{th}})} \nonumber \\ &= -(s v_{\mathrm{th}})^{(2-\alpha)/(\alpha + 1)} \frac{1}{\xi} \frac{\operatorname{Im} \int^{\infty}_{0} dy \, e^{-i \xi y} y^{\alpha-1} \phi(y)}{ \operatorname{Re} \int^{\infty}_{0} dy \, e^{-i \xi y} \phi(y) }.
\end{align}
Since $\hat{\Gamma}^{k}$ is positive when $\xi \gg 1$ and negative when $\xi \ll 1$, viz., \eqref{asymptotics_k_flux}, there is always an order-unity $\xi_{\alpha}$ at which $\hat{\Gamma}^{k} = 0$. Unlike the Batchelor case \eqref{flux_alpha2_ratio}, the ratio of fluxes \eqref{flux_alpha_ratio} is not a function solely of $\xi$. Therefore, \eqref{flux_alpha_ratio} implies that along curves of constant $\xi \sim 1 \neq \xi_{\alpha}$, the flux is always diffusion-dominated as $s v_{\mathrm{th}} \rightarrow \infty$. Furthermore, in the asymptotic region $\xi \gg 1$ (which for $\alpha = 2$ was the phase-mixing-dominated region), we show in Appendix \ref{phases_k_flux} that, as $\alpha$ gets smaller, the region where the flux is asymptotically phase-mixing-dominated shrinks. In fact, for $\alpha < 1/2$, there is no asymptotic region at all where the flux is phase-mixing-dominated (except along the curve $\xi = \xi_{\alpha}$).

As an example, we plot the (nondimensionalized) vector flux for $\alpha = 1$ in Fig.~\ref{fig:a1_flux_ks_alpha2_alpha1}(b) (explicit expressions for the components of the flux in this case can be found in Appendix \ref{alpha1_calc_section}). In this case, apart from along the curve $\xi = \xi_1$ where $\hat{\Gamma}^{k} = 0$, the flux is phase-mixing-dominated only in the region $|s v_{\mathrm{th}}| \lesssim 1$ (irrespective of~$k$), which in Fig.~\ref{fig:a1_flux_ks_alpha2_alpha1}(b) we indicate with gray lines.

These results do not mean that there is no effective phase mixing for these cases. The ratio \eqref{ratio_time_scale} of nonlinear and linear time scales gives the relative local transport of $\delta C_2$ in $s$ versus $k$ space. While the non-local transport of $\delta C_2$ in $k$ space dominates over its flux in $s$, locally, the phase-mixing time \eqref{p_time} of $\hat{F}$ is still shorter than the diffusion time \eqref{d_time} when $\xi \gg 1$. This dominant local phase mixing is what sets up the lowest-order, constant-flux-in-$s$ spectrum in the $\xi \gg 1$ region, viz., \eqref{linear_constant_flux} and~\eqref{asymptotics_spec_twiddle}.

The fluxes also obey critical balance. It is straightforward to show, using \eqref{asymptotics_k_flux} and \eqref{asymptotics_s_flux} with calculations analogous to \eqref{spec_1D_k_integral_twiddle} and \eqref{spec_1D_s_integral_twiddle}, that the 1-D fluxes \eqref{flux_k_alpha} and~\eqref{s_flux_constant} are dominated by contributions from the critical-balance region \eqref{CB} ($\xi \sim 1$). Even though the 2-D $s$ flux is subdominant to the $k$ flux, the fact that $L \int dk \, \hat{\Gamma}^{s}$ is constant in the $s$ inertial range implies that phase mixing still provides an effective route to dissipation scales in velocity space.

\subsection{Shell-averaged flux} \label{sec_SA_flux}

To understand the net effect of having both phase-mixing modes that propagate from low to high $|s|$ and phase-unmixing modes that propagate from high to low~$|s|$, it is useful to consider the flux shell-averaged in~$k$,
\begin{align} \label{Gamma_shell}
    \mathbf{\bar{\Gamma}} \equiv \left(\hat{\Gamma}^k(k,s)-\hat{\Gamma}^k(-k,s), \hat{\Gamma}^s(k,s) + \hat{\Gamma}^s(-k,s) \right).
\end{align}
Note that in 1D, shell averaging amounts simply to adding together contributions from $+k$ and $-k$. The flux \eqref{Gamma_shell} is defined so that the shell-averaged spectrum $\bar{F} \equiv \hat{F}(k,s) + \hat{F}(-k,s)$ satisfies the equation
\begin{equation}
    \frac{\partial \bar{F}}{\partial t} + \nabla \cdot \mathbf{\bar{\Gamma}} = 2 \, \hat{S} - 2 \nu s^2 \bar{F}.
\end{equation}

\begin{figure}
	\centering
	\includegraphics[width=1\linewidth]{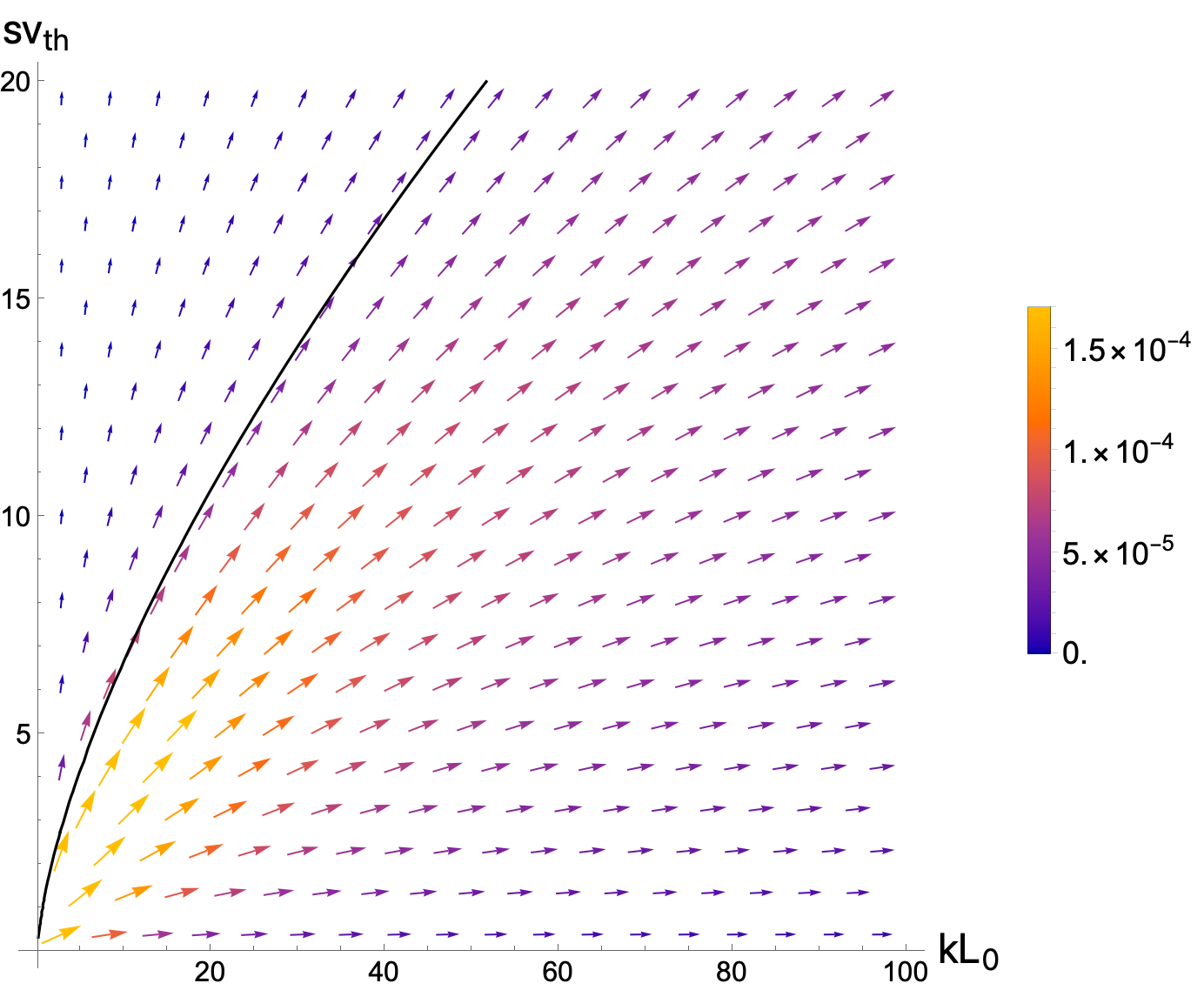}
	\caption{\label{fig:a1_flux_ks_shell_avg} 2-D nondimensionalized shell-averaged (in $k$) flux~\eqref{Gamma_shell} for $\alpha = 1$, with $k$ component $L_0^4 \bar{\Gamma}^{k}(k L_0,s v_{\mathrm{th}})$ and $s$ component $L_0^2 v_{\mathrm{th}}^{6/(\alpha + 1)} \bar{\Gamma}^{s}(k L_0,s v_{\mathrm{th}})$. For emphasis, we specify the magnitude of the flux by both the length and color of the vectors. The black curve is the critical-balance curve~\eqref{CB} for these parameters, $s v_{\mathrm{th}} = (k L_0/\xi_1)^{2/3} $, where $\xi_1 = 1/\sqrt{3}$.}
\end{figure}

While the components of \eqref{Gamma_shell} depend on $\alpha$, their (nondimensionalized) ratio does not (up to a prefactor). Note that, using \eqref{phi_eq} and integrating by parts, \eqref{gamma_k_phi} can be rewritten as
\begin{align} \label{k_flux_in_terms_of_spec}
    &\hat{\Gamma}^{k}(k,s) = 2 \, \varepsilon \, L^{-1} s^{-1} \nonumber \\  &\times \frac{3}{\alpha+1} \operatorname{Re} \left[\phi'(0^{+}) + \xi^2 \int^{\infty}_{0} dy \, e^{-i \xi y} \phi(y) \right],
\end{align}
where the derivative of $\phi$ is taken at $y \rightarrow 0^{+}$. Using \eqref{gamma_s_phi} and \eqref{k_flux_in_terms_of_spec}, we get
\begin{align} \label{flux_SA_ratio}
 \bar{R} &= \frac{L_0^4 \left[\hat{\Gamma}^{k}(k L_0,s v_{\mathrm{th}})-\hat{\Gamma}^{k}(-k L_0,s v_{\mathrm{th}})\right]}{L_0^2 v_{\mathrm{th}}^{6/(\alpha + 1)} \left[\hat{\Gamma}^{s}(k L_0,s v_{\mathrm{th}}) + \hat{\Gamma}^{s}(-k L_0,s v_{\mathrm{th}})\right]} \nonumber \\ &= \frac{3}{\alpha+1} \frac{k L_0}{s v_{\mathrm{th}}}.
\end{align}
Remarkably, this expression is valid everywhere in the $(k,s)$ plane, independent of $\xi$ being small or large. The flux is radial at both large $s v_{\mathrm{th}}$ and large $k L_0$. We plot $\bar{\mathbf{\Gamma}}$ for $\alpha = 1$ in Fig.~\ref{fig:a1_flux_ks_shell_avg}, which clearly exhibits this feature.

Note that the components of the shell-averaged flux are positive-definite. We are only able to show directly that this property holds for $\alpha = 1$; this calculation can be found in Appendix \ref{alpha1_calc_section}. An argument as to why the fluxes are positive-definite for the general case is as follows. Since \eqref{flux_SA_ratio} is positive, the components of \eqref{Gamma_shell} are either both positive or both negative for all $(k,s)$ in the inertial range (no sign reversals are possible, otherwise the flux would not be divergence-free in the inertial range: see \eqref{final_eq_flux_ss}). If the components were negative-definite, there would be a sink at the origin. However, there is a source at the origin, so the shell-averaged fluxes must therefore be positive-definite.

This positive definiteness is important. The circulatory nature of the fluxes in Fig.~\ref{fig:a1_flux_ks_alpha2_alpha1} is due to phase-mixing modes (with $\mathrm{sgn}(ks) = 1$) and phase-unmixing modes (with $\mathrm{sgn}(ks) = -1$) propagating in opposite directions in $s$. Since the $s$ component of \eqref{Gamma_shell} is equal to $k[\hat{F}(k,s) - \hat{F}(-k,s)]$, the shell-averaged flux being positive-definite means that the spectral amplitudes of the phase-mixing modes are greater than those of the phase-unmixing modes. Therefore, by adding together the fluxes of the two modes, we are left with a net flux that points outward to both high $k$ and high $s$ toward the dissipation scales, in agreement with our analysis in Sections \ref{Flux_analysis} and \ref{Flux_analysis_non_local}. 

\section{Conclusion} \label{discussion}

\subsection{Summary}

In this paper, we have presented a solvable model of kinetic plasma turbulence, in which the electric field is decoupled from the particle distribution function and taken to be an externally imposed Gaussian field, white-noise in time and power-law in $k$ space.

The effect of this stochastic electric field on the mean distribution function is diffusion in velocity space, often referred to as stochastic heating \cite{sturrock1966stochastic,chandran2010perpendicular,verscharen2019multi,cerri2021stochastic}. The resulting energization of particles is a collisionless process. Indeed, the heating rate is set by the turbulent collisionality and is independent of $\nu$ [see \eqref{KE_evol}]. However, the irreversibility of stochastic heating hinges on the presence of collisions. As $\langle f \rangle$ heats, $\delta \! f$ fluctuations are excited, transferring the minus `entropy' $C_{2,0} = (1/L)\int dx dv \, \langle f \rangle^2/2$ into $\delta C_2 = (1/L)\int dx dv \, \langle \delta \! f^2 \rangle/2$, which then cascades to small scales in both position and velocity space simultaneously. This cascade is then cut off by collisions at fine phase-space scales, thereby rendering the heating irreversible. The irreversibility of stochastic heating is therefore enabled by the entropy cascade.

We have analyzed this cascade in the Fourier-transformed $(k,s)$ space, where an `inertial range' forms to bridge the injection and dissipation scales of $\delta C_2$. Integrated over $k$ or $s$, the flux of $\delta C_2$ is constant in this inertial range. Importantly, there is no collisional dissipation at scales much larger than the dissipation scale $(\ell_{\nu},u_{\nu})$ [see \eqref{kolmogorov_scale} and \eqref{u_nu_k_nu}], which tends to $(0,0)$ as the collision frequency does.

In the 2-D $(k,s)$ space, the Fourier spectrum of $\delta C_2$ has a self-similar profile, with power-law asymptotics at high $k$ and $s$, respectively. We find that these asymptotic scalings can be deduced by a phenomenological theory whose governing principle is that the cascade satisfies a critical balance in phase space between the time scales of linear phase mixing and turbulent diffusion. Because there is nothing in our phenomenological theory that is unique to 1D-1V, we also expect these ideas to apply in 2D-2V and 3D-3V. While the one-dimensional $|k|$ and $|s|$ spectra \eqref{spec_1D_k_twiddle} and \eqref{spec_1D_s_twiddle} should be the same, the two-dimensional~$|k|$-$|s|$ spectrum (integrated over angles) will not be the same as~\eqref{asymptotics_spec_twiddle} because of the wavenumber Jacobian being different from unity in higher dimensions.

\subsection{Fast cascade and the effectiveness of phase mixing} \label{C2_capacity}

In a linear system, phase mixing acts as a route to collisional dissipation at every spatial scale \cite{zocco2011reduced,kanekar2015fluctuation}, but in the model presented here, collisional dissipation is only non-negligible below the dissipation scale $\ell_{\nu}$. Following the commonly held intuition that the effect of phase mixing (and Landau damping) on turbulent systems is that it steepens the spectrum (of, e.g., electromagnetic energy) at every scale by an amount set by the Landau-damping rate \cite{hammett1990fluid,hammett1992fluid,dorland1993gyrofluid,podesta2010kinetic,howes2010prescription,passot2015model,zhou2023electron}, one may be led to conclude from our results that phase mixing is less effective in a nonlinear system than in a linear one. However, in fact, phase mixing is even more effective in a nonlinear system. This is because the presence of nonlinearity produces fine structure in position space and thus enhances the rate at which the distribution function develops fine phase-space gradients, reducing the time that it takes collisions to activate,~$\tau_{\nu}$.

To see this, note that in a steadily driven system, as considered in this paper, the total $\delta C_2$ in steady state divided by the injection rate $\varepsilon$ gives a reasonable estimate for $\tau_{\nu}$. For a linear system, restricting ourselves to a single $k$ and noting that the spectrum is flat in $s$ \footnote{To see this, consider the inertial-range equation for the spectrum \eqref{final_eq_inertial_range}. In the absence of nonlinearity, this equation is just $k \, \partial_s \hat{F} = 0$, whose solution is just $\hat{F} = \textrm{const}$.} up to a collisional cutoff $s_{\nu} \propto \nu^{-1/3}$ given by the balance~\eqref{bal_fs_coll}~\cite{kanekar2015fluctuation,zocco2011reduced}, we find \cite{su1968collisional}
\begin{equation} \label{tau_nu_L}
   \tau_{\nu} \sim \frac{\delta C_2}{\varepsilon} \propto \int^{s_{\nu}} ds \propto \nu^{-1/3}. 
\end{equation}
For smaller and smaller $\nu$, it takes longer and longer for collisions to dissipate the velocity-space structure of the distribution function, and in the collisionless limit, the amount of $\delta C_2$ stored in phase space diverges \cite{kanekar2015fluctuation}. This is consistent with a constant cascade time, set by the phase-mixing time, viz., $\tau_c \sim (k \, v_{\mathrm{th}})^{-1}$ (with $k$ fixed).

In the presence of nonlinearity, when $\alpha = 2$, the 1-D spectra \eqref{spec_1D_k_twiddle} and \eqref{spec_1D_s_twiddle} scale like $k^{-1}$ and $s^{-1}$, respectively~\cite{adkins2018solvable}, the former scaling in agreement with the classical Batchelor \cite{batchelor1959small} spectrum of a passive scalar advected by a single-scale flow. Integrating the 1-D spectrum up to the collisional cutoff \eqref{kolmogorov_scale} gives
\begin{equation} \label{delta_C2_Batchelor}
    \tau_{\nu} \sim \frac{\delta C_2}{\varepsilon} \propto \int^{1/\ell_{\nu}} dk \, k^{-1} \propto | \log \nu |.
\end{equation}
Although formally also divergent as $\nu \rightarrow 0^{+}$, \eqref{delta_C2_Batchelor} is asymptotically shorter than \eqref{tau_nu_L}. In this case, the cascade time is also constant, viz., $\tau_{\mathrm{c}} \sim \kappa^{-1/3}$, as can be seen in \eqref{tau_c}, even though many $k$'s are involved.

When $\alpha < 2$, the phase-space cascade is even more efficient. From \eqref{tau_c}, we have that $\tau_{\mathrm{c}}$ goes to zero as $k$ and $s$ go to infinity, so fluctuations `turn over' faster to finer phase-space scales the deeper they are in the inertial range (compared to the constant cascade times in the linear and Batchelor regimes). As a result, the 1-D spectra \eqref{spec_1D_k_twiddle} and \eqref{spec_1D_s_twiddle} are steeper than $k^{-1}$ and $s^{-1}$, so the total steady-state $\delta C_2$ and~$\tau_{\nu}$ are independent of~$\nu$.

Thus, the presence of nonlinearity reduces the collision time from a (negative) fractional power of $\nu$ \eqref{tau_nu_L} in the linear regime to a time scale that is only logarithmic with~$\nu$ \eqref{delta_C2_Batchelor} when $\alpha = 2$, or one that is independent of $\nu$ when $\alpha < 2$. To interpret this shortening of $\tau_{\nu}$ in the nonlinear regime, note that linear phase mixing still processes $\delta C_2$ from injection to dissipation scales in the inertial range, but so does, simultaneously, nonlinear mode coupling, in a critically-balanced fashion. The net result is fast dissipation, but only at wavenumbers $ k \gtrsim k_{\nu} \sim 1/\ell_{\nu}$ and $ s \gtrsim s_{\nu} \sim 1/u_{\nu}$, which, by construction, satisfy the critical-balance condition \eqref{CB}. The nonlinear cascade ensures that all the injected $\delta C_2$ flux at large phase-space scales is rapidly dissipated at small scales via collisions, no matter how small is $\nu$. This reduction of $\tau_{\nu}$ is also an \textit{a posteriori} justification of the assumption \eqref{deltaC2balance_ss} of there being a separation of time scales between the time that it takes $\delta C_2$ to reach quasi-steady state and the time that it takes for the injection rate $\varepsilon$ to decay due to the stochastic heating of~$\langle f \rangle$ \footnote{Formally, this assumption is not valid in the Batchelor regime in the collisionless limit, since the total $\delta C_2$ stored in phase space and collision time $\tau_{\nu}$ diverge, viz.,~\eqref{delta_C2_Batchelor}. However, this divergence is only logarithmic, so one may expect quasi-steady states in the Batchelor regime to be possible when $\nu$ is finite.}.

\subsection{Implications and outlook}

As discussed in Section \ref{C2_capacity}, these results provide a conceptual understanding of the role of phase mixing in turbulent plasmas. Recent theoretical \cite{schekochihin2016phase,adkins2018solvable} and numerical \cite{parker2016suppression,meyrand2019fluidization} studies have suggested a statistical `fluidization' of turbulent, collisionless plasmas by stochastic plasma echoes suppressing phase mixing in the (spatial) inertial range. This might have seemed to be at odds with Landau damping \cite{landau1946vibrations} clearly being identified in turbulent settings numerically \cite{tenbarge2013current,navarro2016structure,klein2017diagnosing,howes2018spatially,nattila2022heating,zhou2023electron} in several works, as well in Magnetospheric Multiscale (MMS) mission observations of the turbulent plasma in the Earth's magnetosheath \cite{chen2019evidence,afshari2021importance}. While our model is quite simplified and does not include self-consistent electric fields and hence Landau damping, insofar as Landau damping and phase mixing are intimately related processes, our results indicate that these two seemingly contradictory sets of results can in fact be compatible.

This work also has implications for the relaxation of mean distribution functions in nearly collisionless plasmas. The existence of the entropy cascade implies that collisions will always dissipate fine-scale structure in the distribution function, even when $\nu$ is vanishingly small, but it does not necessarily imply that the rate by which the distribution function relaxes toward a Maxwellian is enhanced. This is clear in our model from the fact that, whereas $\delta \! f$ develops sharp phase-space gradients, $\langle f \rangle$ does not, so Coulomb collisions are never activated for it. Mean distributions in space plasmas can be highly non-Maxwellian, e.g., in the solar wind \cite{marsch2006kinetic,verscharen2019multi,martinovic2020enhancement}. Developing a theoretical formalism to predict the form of such non-equilibrium distribution functions is an outstanding problem. A direct consequence of entropy cascades is that theories of relaxation that assume phase-volume conservation \cite{lynden1967statistical,kadomtsev1970collisionless,chavanis2021kinetic,ewart2022,ewart2023nonthermal} may not apply to nearly collisionless plasmas that are strongly turbulent. An alternative approach is to examine how the turbulent phase-space correlations of $\delta \! f$ drive the evolution of the mean distribution function \cite{dupree1972theory,diamond2010modern}.

There is much opportunity to understand phase-space entropy cascades in nearly collisionless plasmas better, with theory, numerical simulations, laboratory experiments, and spacecraft data. With regards to the latter two, the works \cite{kawamori2013verification,kawamori2022evidence,servidio2017magnetospheric,wu2023ion} suggest that measuring entropy cascades in real plasmas is a realizable endeavor just beginning to be possible.

M.L.N is grateful to I. Abel, T. Adkins, M. Allen, W. Clarke, S. Cowley, P. Dellar, J.-B. Fouvry, M. Kunz, R. Meyrand, F. Rincon, and J. Squire for helpful discussions related to this work. W.S would like to thank A. Bhattacharjee and H. Weitzner for useful discussions. We are also grateful to two anonymous referees, whose feedback improved the manuscript. M.L.N was supported by a Clarendon Scholarship. R.J.E was supported by a UK EPSRC studentship. W.S. was supported by a grant from the Simons Foundation/SFARI (560651, AB). The work of A.A.S. was supported in part by UK EPSRC (grant EP/R034737/1), STFC (grant ST/W000903/1), and the Simons Foundation via a Simons Investigator award. W.D.D. and M.L.N were supported by the US Department of Energy through grant DEFG0293ER54197 and Scientific Discovery Through Advanced Computing (SciDAC) grant UTA18000275.

\appendix

\section{Invariants alternative to \texorpdfstring{$C_2$}{C2}} \label{heating_sec}

In this appendix, we discuss invariants of the Vlasov equation alternative to $C_2$. It is straightforward to show that \eqref{Vlasov} conserves an infinite number of invariants, so-called Casimir invariants \cite{ye1992action}. Indeed, focusing on invariants defined in the averaged sense, for any smooth $g(f)$, the functional
\begin{equation} \label{CasimirG}
    G[f] = \frac{1}{L} \iint dx dv \, \langle g(f) \rangle
\end{equation}
satisfies, using \eqref{approx_coll_op_delta_f} for the collision operator,
\begin{align}
    \frac{d G}{d t} &= \frac{1}{L} \iint dx dv \,  \langle g'(f) C[f] \rangle \nonumber \\ &= - \frac{\nu}{L} \iint dx dv \,  \left\langle g''(f) \left(\frac{\partial \langle f \rangle}{\partial v} + \frac{\partial \delta \! f}{\partial v} \right) \frac{\partial \delta \! f}{\partial v}\right \rangle \nonumber \\ & \simeq - \frac{\nu}{L} \iint dx dv \, \left \langle g''(f) \left| \frac{\partial \delta \! f}{\partial v} \right|^{2} \right \rangle,
\end{align}
where the term with two derivatives on $\delta \! f$ dominates when $\nu \rightarrow 0^{+}$. Note that while in this paper $\langle ... \rangle$ denotes ensemble averaging with respect to the stochastic electric field, the arguments in this section hold for any averaging procedure for which one can decompose $f$ = $\langle f \rangle + \delta \! f$, where $\langle \delta \! f \rangle = 0$.

When $\nu = 0$, every $G[f]$ is formally conserved. Furthermore, if $g(f)$ is convex, i.e., $g''(f) \geq 0$ everywhere, then the corresponding Casimirs are negative-definitely dissipated by collisions. We label Casimirs with positive-definite time evolution (which are minus the convex functionals of $f$) as `generalized entropies' (cf. entropy functions in hyperbolic partial differential equations \cite{godlewski2013numerical}). This set includes $-C_2$, as well as the traditional entropy~${S = - \iint dx dv \, f \log f}$.

We define the relative entropy as
\begin{equation} \label{R_entropy}
    R[f] = G[f] - G[\langle f \rangle],
\end{equation}
which has the budget
\begin{equation} \label{g_budget}
    \frac{d R[f]}{dt} + \frac{d G[\langle f \rangle]}{dt} = - \frac{\nu}{L} \iint dx dv \, \left\langle g''(f) \left| \frac{\partial \delta \! f}{\partial v} \right|^{2} \right \rangle.
\end{equation}
Using \eqref{vlasov_avg_diff}, $G[\langle f \rangle]$ satisfies
\begin{equation}
    \frac{d G[\langle f \rangle]}{dt} = -D_0 \int dv \, g''(\langle f \rangle) \left| \frac{\partial \langle f \rangle}{\partial v} \right|^{2},
\end{equation}
which is negative-definite if $g(f)$ is convex. Therefore, we can interpret \eqref{g_budget} analogously to \eqref{deltaC2balance}. In the absence of collisions, as $G[\langle f \rangle]$ decreases via stochastic heating, $R[f]$ increases to maintain the $G[f]$ balance. Once $\delta \! f$ has developed sharp enough gradients, collisions dissipate the total $G[f]$.

We have shown that $C_2$ is anomalously dissipated as $\nu \rightarrow 0^{+}$ and is cascaded, i.e., exhibits an inertial range unaffected directly by collisions or forcing. In principle, invariants other than $C_2$ can also have these properties. A system mathematically similar to \eqref{Vlasov} in which this happens is the advection-diffusion equation for a scalar advected by a turbulent flow: \cite{falkovich1994turbulence} showed that the family of invariants $\int d\textbf{x} \, \theta^{2n}$, where $\theta$ is a scalar field and $n$ is a positive integer, satisfy constant-flux cascades; in contrast, for a passive scalar advected by a smooth, chaotic flow (Batchelor regime), only the quadratic invariant ($n = 1$) is cascaded. This is because for higher-order invariants,  logarithmic correlations of the passive scalar give rise to injection of those invariants by the source at all scales, preventing the formation of an inertial range.  

It is likely that a similar situation happens in the Vlasov-Kraichan model between the cases $\alpha < 2$ and $\alpha = 2$, but such a calculation is beyond the scope of this work. If it were true, then for $\alpha < 2$, the cascade of $C_2$ does not necessarily hold deeper physical meaning than the cascade of any other convex functional of $f$.

However, we still believe that $C_2$ is a particularly useful quantity in kinetic plasma turbulence. Because it is quadratic in $f$, $C_2$ is the only invariant (up to weight functions in the integrand of \eqref{CasimirG}) that satisfies

\begin{equation}
     G[\langle f \rangle + \delta \! f] = G[\langle f \rangle] + G[\delta \! f],
\end{equation}
and so the relative entropy \eqref{R_entropy} is a function solely of $\delta \! f$. This property is useful for conceptualizing the budget \eqref{g_budget} as a transfer of entropy between $\langle f \rangle$ and $\delta \! f$, as any other Casimir invariant involving higher powers of $f$ will necessarily involve cross terms containing both $\langle f \rangle$ and $\delta \! f$. Furthermore, $C_2$ is the only invariant that lends itself to a simple Fourier analysis. For these reasons, in this paper, we have chosen to analyze phase-space turbulence using $C_2$ exclusively.

\section{Derivation of the Fourier-spectrum equation} \label{error_estimate_sec} 

In this appendix, we give a detailed derivation of \eqref{spec_eq_FL_noflux}.

Multiplying \eqref{fhat_eq} by $\delta \hat{f}^{*}(k,s)$, adding to the resulting equation its complex conjugate, and then ensemble averaging gives
\begin{align} \label{vlasov_spec_unsimp}
    \frac{\partial \hat{F}}{\partial t} + k\frac{\partial \hat{F}}{\partial s} + s \, \mathrm{Im} \sum_p \, \left \langle \hat{E}(p) \delta \hat{f}(k-p) \delta \hat{f}^* (k) \right \rangle  \nonumber \\ = \hat{S} - 2 \nu s^2 \hat{F}.
\end{align}
Note that the second term in the Fourier sum in \eqref{fhat_eq} vanishes under multiplication by $\delta \hat{f}^{*}(k,s)$ and ensemble averaging. The source term $\hat{S}$, defined in \eqref{source_sk}, comes from the second term on the right-hand side of \eqref{fhat_eq}. It can be found via application of the Furutsu-Novikov theorem \eqref{NF} and using the fact that \eqref{electric_field_correlator} implies
\begin{equation} \label{E_corr}
\langle \hat{E}(k,t) \hat{E}(k',t') \rangle = 2  \,\hat{D}(k) \delta_{k,-k'} \delta(t-t'),
\end{equation}
where $\delta_{k,-k'}$ is the Kronecker delta.

\subsection{Derivation of \texorpdfstring{\eqref{C2_Fhat_balance}}{(31)}}

The $\delta C_2$ budget \eqref{C2_Fhat_balance} in terms of $\hat{F}$ can be found by taking the time derivative of \eqref{Planch} and using \eqref{vlasov_spec_unsimp}. Assuming the spectrum goes to zero at $s \rightarrow \pm \infty$, the free-streaming term vanishes by integration over $s$. For the nonlinear term, note that the summand in
\begin{align}
    & \mathrm{Im} \sum_{k,p} \, \left \langle \hat{E}(p) \delta \hat{f}(k-p,s) \delta \hat{f}^* (k,s) \right \rangle,
\end{align}
after taking $p \rightarrow -p$ and $k \rightarrow k-p$, and applying reality conditions on the electric field, is equal to its own complex conjugate. Therefore, its imaginary part vanishes. What is left is injection by the source and dissipation by collisions, viz., \eqref{C2_Fhat_balance}.

\subsection{Derivation of \texorpdfstring{\eqref{spec_eq_FL_noflux}}{(32)}}

We now close the triple correlator in \eqref{vlasov_spec_unsimp}. Using \eqref{NF}, we have
\begin{align} \label{FN_2}
    &\left\langle \hat{E}(p) \delta \hat{f}(k-p)  \delta \hat{f}^* (k) \right\rangle \nonumber = \int dt' \sum_{p'} \left\langle \hat{E}(p,t) \hat{E}(p',t') \right\rangle \\  & \times \left \langle \frac{ \delta \left[ \delta \hat{f}(k-p,t) \delta \hat{f}^* (k,t) \right] }{\delta \hat{E}(p',t')}  \right \rangle.
\end{align}
As in Section \ref{SA}, the functional derivative can be computed by formally integrating the relevant evolution equation. Using \eqref{fhat_eq}, we can write
\begin{align}
    & \delta \hat{f}(k-p) \delta \hat{f}^*(k) \nonumber \\ = & \int^{t} dt'' \, \bigg\{is \sum_{p''}\bigg[\hat{E}(p'') \delta \hat{f}(k-p-p'') \delta \hat{f}^*(k) \nonumber \\ & -\hat{E}(-p'') \delta \hat{f}^*(k-p'') \delta \hat{f}(k-p)\bigg] + (...) \bigg\},
\end{align}
where $(...)$ represents terms that will vanish after we take the functional derivative. Therefore, we get
\begin{align}
&\left \langle \frac{ \delta \left[ \delta \hat{f}(k-p,t) \delta \hat{f}^* (k,t) \right] }{\delta \hat{E}(p',t')}  \right \rangle   \nonumber  \\  &=   is \bigg[\left\langle  \delta \hat{f}(k-p-p',t') \delta \hat{f}^*(k,t') \right\rangle \nonumber \\ &- \left\langle \delta \hat{f}^*(k+p',t') \delta \hat{f}(k-p,t') \right\rangle  \bigg]H(t-t'),
\end{align}
where $H$ is the Heaviside step function, defined with the convention that $H(0) = 1/2$. Combining this expression with \eqref{E_corr} and \eqref{FN_2} yields
\begin{align} \label{triple_correlator}
    \langle \hat{E}(p) \delta \hat{f}(k-p) \delta \hat{f}^* (k) \rangle = 2 i s \, \hat{D}(p) \left[\hat{F}(k)-\hat{F}(k-p)\right].
\end{align}
Using this expression in \eqref{vlasov_spec_unsimp}, we get
\begin{align} \label{spec_eq_conv}
    \frac{\partial \hat{F}}{\partial t} + k\frac{\partial \hat{F}}{\partial s} + 2 s^2 \, \sum_p \, \hat{D}(p) \left[\hat{F}(k)-\hat{F}(k-p)  \right]  \nonumber \\ = \hat{S} - 2 \nu s^2 \hat{F}.
\end{align}

\subsection{The case of \texorpdfstring{$\alpha = 2$}{alpha = 2}: Batchelor limit} \label{app_Batch}

Let us simplify the nonlinear term in \eqref{spec_eq_conv} further. We start with the case of $\alpha = 2$. In the limit $k L_{E} \gg 1$,  we suppose (and check \textit{a posteriori}) that \eqref{hatD} is sufficiently steep in wavenumbers that we can consider $k \gg p$ and Taylor-expand the summand of the wavenumber sum in~\eqref{spec_eq_conv}:
\begin{equation} \label{Batch_approx}
    \hat{F}(k)-\hat{F}(k-p) \simeq -p \frac{\partial \hat{F}}{\partial k} + \frac{1}{2}p^2 \frac{\partial^2 \hat{F}}{\partial k^2}.
\end{equation}
This `Batchelor approximation' was first used for the problem of passive-scalar mixing in fluids \cite{batchelor1959small,kraichnan1974convection} and amounts to approximating the electric field as effectively single-scale. Substituting \eqref{Batch_approx} back into the sum in \eqref{spec_eq_conv}, the first term vanishes because it is odd in $p$, and we are left with
\begin{equation} \label{Batch_approx2}
    \sum_p \, \hat{D}(p) \left[\hat{F}(k)-\hat{F}(k-p) \right] \simeq D_2 \frac{\partial^2 \hat{F}}{\partial k^2},
\end{equation}
where
\begin{equation} \label{kappa_batchelor}
    D_2 = \frac{1}{2}\sum_p p^2 \hat{D}(p) =  \frac{D}{2} \sum_p \frac{p^2 e^{-(\eta p)^2}}{(p^2+L_{E}^{-2})^{3/2}}.
\end{equation}
In the limit $\eta \rightarrow 0^{+}$, \eqref{kappa_batchelor} is logarithmically divergent, being $\propto~\log \left( L_{E}/ \eta \right)$; without a small-scale cutoff, the approximation \eqref{Batch_approx2} is invalid. This is because the~$k^{-3}$ spectrum in \eqref{hatD} is only marginally in the Batchelor regime \cite{batchelor1959small, adkins2018solvable}. The Batchelor limit generically applies when the electric field is spatially smooth, corresponding to a rapidly decaying spectrum $\hat{D}(k)$. We choose the particular form of $\hat{D}(k)$ in \eqref{hatD} in order to match onto the Batchelor limit and fractional cases with one functional form of the correlation function, but we could just as well have picked a steeper \eqref{hatD} for $\alpha = 2$ that would not have required a small-scale cutoff. Therefore, without loss of generality, we keep \eqref{kappa_batchelor} without modification.

\subsection{The \texorpdfstring{$\alpha < 2$}{alpha < 2} cases: representation in terms of the fractional Laplacian} \label{app_frac_Lap_deriv}

For $\alpha < 2$, it is convenient to manipulate the nonlinear term in \eqref{spec_eq_conv} in position space and then Fourier transform back to $k$ space. We start by noting that
\begin{align}
   \sum_k e^{ikr} \sum_p \hat{D}(p) &\left[\hat{F}(k)-\hat{F}(k-p)  \right] \nonumber \\ = &\left[D(0)-D(r)\right]F(r).
\end{align}
We now take the limits $\eta \rightarrow 0^{+}$ and $k L_{E} \gg 1$. When $\eta = 0$, note that \eqref{Dr} is the kernel of the Bessel potential \cite{aronszajn1961theory,eyink2000self}, and can therefore be written as
\begin{equation} \label{Dr_Bessel}
    D(r) = \frac{L D}{2 \pi} \frac{2^{1-\alpha/2}\sqrt{\pi}L^{\alpha}_{E}}{\Gamma\left(\frac{\alpha+1}{2}\right)} \left(\frac{r}{L_{E}}\right)^{\alpha/2} \, K_{\alpha/2}\left(\frac{r}{L_{E}}\right),
\end{equation}
where $K$ is the modified Bessel function of the second kind, and $\Gamma$ is the Gamma function. For $\alpha < 2$ and $r/ L_{E} \ll 1$, \eqref{Dr_Bessel} has a series expansion
\begin{equation} \label{Dr_series}
    D(r) = D_0 - D_{\alpha} r^{\alpha} + \mathcal{O}\left(\frac{r^2}{L^{2-\alpha}_{E}}\right),
\end{equation}
where
\begin{equation} \label{D_0}
    D_0 = \frac{L D}{2 \pi} \frac{\Gamma\left(\frac{\alpha}{2}\right)\sqrt{\pi}}{\Gamma\left(\frac{\alpha+1}{2}\right)} L^{\alpha}_{E},
\end{equation}
\begin{equation} \label{D_alpha}
    D_{\alpha} = \frac{L D}{2 \pi} \frac{ \sqrt{\pi} \, \left|\Gamma\left(\frac{-\alpha}{2}\right)\right|}{ 2^{\alpha} \, \Gamma\left(\frac{\alpha+1}{2}\right) }.
\end{equation}

We now Fourier transform back to $k$ space. To lowest order in $(r / L_E)^{2-\alpha} \ll 1$ (equivalently, $(k L_E)^{2-\alpha} \gg 1$), the $r^{\alpha}$ term is dominant over the $r^2$ term, which gives
\begin{align} \label{frac_Deriv}
    &\frac{1}{L} \int dr e^{-ikr} \left[D(0)-D(r)\right]F(r) \nonumber  \\ & \simeq \frac{1}{L} \int dr e^{-ikr} D_{\alpha} |r|^{\alpha} F(r) = D_{\alpha} (-\Delta_k)^{\alpha/2} \hat{F}(k).  
\end{align}
Here $(-\Delta_k)^{\alpha/2}$ is a fractional Laplacian \cite{2012hitchhiker,pozrikidis2018fractional,lischke2020fractional} of order $\alpha/2$, in $k$ space, viz.,
\begin{equation} \label{def_frac_lap}
    (-\Delta_k)^{\alpha/2} \hat{F}(k) = c_{\alpha} \,  \text{p.v.} \int_{-\infty}^{+\infty}dp \, \frac{\hat{F}(k)-\hat{F}(k-p)}{|p|^{\alpha+1}},
\end{equation}
where
\begin{equation} \label{calpha}
        c_{\alpha} = \frac{2^{\alpha}\Gamma(\frac{\alpha+1}{2})}{\sqrt{\pi}|\Gamma\left(\frac{-\alpha}{2}\right)|},
\end{equation}
and $\text{p.v.}$ means that the integral is defined in the principal-value sense.

Thus, using \eqref{frac_Deriv} and \eqref{Batch_approx2}, we have that \eqref{spec_eq_conv} becomes \eqref{spec_eq_FL_noflux}, where
\begin{equation} \label{kappa}
    \kappa = \begin{cases} 
       2 D_{\alpha},  \qquad  & \alpha < 2, \\
       2 D_2,  \qquad & \alpha = 2,
   \end{cases}
\end{equation}
with $D_{\alpha}$ given by \eqref{D_alpha} and $D_2$ given by \eqref{kappa_batchelor}.

\subsection{Derivation of \texorpdfstring{\eqref{gamma_k_nonlocal}}{(104)} } \label{derivation_gamma_k_nonlocal}
Here, we derive the form \eqref{gamma_k_nonlocal} of the non-local $k$ flux in the $\alpha < 2$ cases, as analyzed in Section \ref{Flux_analysis_non_local}.

We first rewrite \eqref{def_frac_lap} as an integral over positive $p$:
\begin{align} \label{flux_deriv_pos_p}
(-\Delta_k)^{\alpha/2} & \hat{F}(k) = c_{\alpha} \,  \text{p.v.} \int_{-\infty}^{+\infty}dp \, \frac{\hat{F}(k)-\hat{F}(k-p)}{|p|^{\alpha+1}} \nonumber \\ &= c_{\alpha} \int_{0}^{\infty}dp \, \frac{2 \hat{F}(k) - \hat{F}(k-p) - \hat{F}(k+p) }{p^{\alpha + 1}}.
\end{align}
In \eqref{flux_deriv_pos_p}, we can write $p^{-(\alpha + 1)} = -(1/\alpha) \partial_p p^{-\alpha}$ and integrate by parts. The boundary terms vanish (assuming $\hat{F}$ vanishes at infinity), and we are left with
\begin{align} \label{flux_deriv_2}
(-\Delta_k)^{\alpha/2}  \hat{F}(k) &= -\frac{c_{\alpha}}{\alpha} \int_{0}^{\infty}dp \, \frac{\frac{\partial \hat{F}}{\partial p}(k-p) + \frac{\partial \hat{F}}{\partial p}(k+p)}{p^{\alpha}} \nonumber \\ &= \frac{\partial}{\partial k} \frac{c_{\alpha}}{\alpha} \int_{0}^{\infty}dp \, \frac{\hat{F}(k-p) - \hat{F}(k+p)}{p^{\alpha}}.
\end{align}
Combining \eqref{flux_deriv_2} and the definition $\partial_k \hat{\Gamma}^{k} = \kappa s^2 (-\Delta_k)^{\alpha/2} \hat{F} $ yields \eqref{gamma_k_nonlocal}.

\section{Detailed calculations for Section \texorpdfstring{\ref{const_flux}}{IV A}} \label{branch_cut_calc_section}

In this appendix, we derive the general $\alpha < 2$ expressions for $\hat{g}(k)$ \eqref{gk_branch} and the $k$ flux integrated over $s$ \eqref{flux_k_alpha}.

\subsection{Derivation of \texorpdfstring{\eqref{gk_branch}}{(63)} and \texorpdfstring{\eqref{dissp_k_alpha}}{(70)} } \label{det_calc_subsec1}

\begin{figure}
	\centering
	\includegraphics[width=1\linewidth]{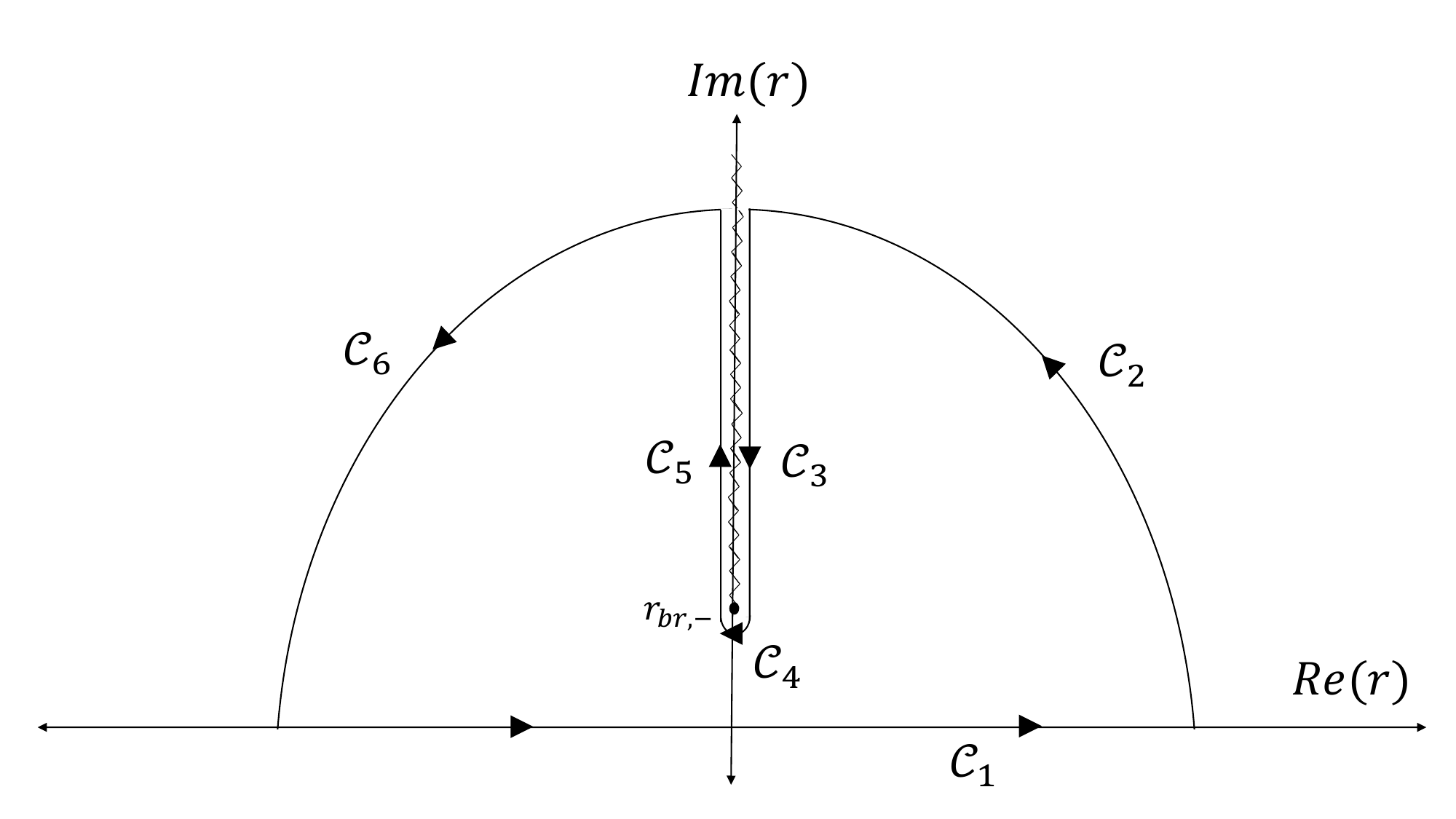}
	\caption{\label{fig:contour_branch_cut} Contour used in the derivation of \eqref{gk_branch}.}
\end{figure}

We start with $\hat{g}(k)$, as given in \eqref{g_k}. As discussed in Section \ref{const_flux}, it is useful to convert this integral into one where the exponential in the integrand is decaying rather than oscillatory. To do this, we perform an auxiliary expansion, taking $|r|^{\alpha} \rightarrow (r^2+\delta^2)^{\alpha/2}$ and then $\delta \rightarrow 0$ in the integral \eqref{g_k}:
\begin{equation}
    \hat{g}(k) = \frac{\varepsilon}{2 \pi \kappa L} \lim_{\delta \to 0} \,  \mathcal{I}_{\mathcal{C}_1},
\end{equation}
where
\begin{equation} \label{IC1}
    \mathcal{I}_{\mathcal{C}_1} = \int_{\mathcal{C}_1} dr \, \frac{e^{-ikr}}{(r^2+\delta^2)^{\alpha/2} + \ell^{\alpha}_{\nu}},
\end{equation}
and $\mathcal{C}_1$ is a contour running along the real line. We choose the branch cuts so that
\begin{align} 
    (r^2+\delta^2)^{\alpha/2} & = (r+i\delta)^{\alpha/2}(r-i\delta)^{\alpha/2} \nonumber \\ & = r^{\alpha/2}_{+}e^{i \alpha \theta_{+}/2} r^{\alpha/2}_{-}e^{i \alpha \theta_{-}/2},
\end{align}
where $r_{\pm} = |r \pm i \delta|$, $\theta_{+} \in [-\pi/2,3 \pi/2 ]$, and  $\theta_{-} \in~[-3 \pi/2,\pi/2]$. Note the branch points are at $r_{\textrm{br}, \pm} = \mp i \delta$. We first treat the case $k < 0$. We close $\mathcal{C}_1$ by making a semi-circle in the upper half of the complex plane and going around the branch cut with the branch point $r_{\textrm{br}, -}$, as depicted in Fig.~\ref{fig:contour_branch_cut}. By Cauchy's residue theorem, since the integrand in \eqref{IC1} has no poles, $ \mathcal{I}_{\mathcal{C}_1} + \mathcal{I}_{\mathcal{C}_2} + \mathcal{I}_{\mathcal{C}_3} + \mathcal{I}_{\mathcal{C}_4} + \mathcal{I}_{\mathcal{C}_5} + \mathcal{I}_{\mathcal{C}_6} = 0$. When the radius $R$ of the contours $\mathcal{C}_2$ and $\mathcal{C}_6$ is large, the integrands along these contours are $ \propto e^{-|k| \, R \sin \theta} \, R^{1-\alpha}$ (here $\theta$ is the angle of the contour with respect to the positive real axis). Thus, $\mathcal{I}_{\mathcal{C}_2}$ and $\mathcal{I}_{\mathcal{C}_6}$ tend to zero as $R \rightarrow \infty$. Because the integrand is finite at $r = r_{\textrm{br}, -}$, $\mathcal{I}_{\mathcal{C}_4}$ tends to zero as the radius of $\mathcal{C}_4$ shrinks to zero. Therefore, $\mathcal{I}_{\mathcal{C}_1} = - \mathcal{I}_{\mathcal{C}_3} - \mathcal{I}_{\mathcal{C}_5}$. Noting that along $\mathcal{C}_3$, $\theta_{\pm} = \pi/2$, while along $\mathcal{C}_5$, $\theta_{+} = \pi/2$ and $\theta_{-} = -3 \pi/2$, we have that
\begin{align}
&-(\mathcal{I}_{\mathcal{C}_3} + \mathcal{I}_{\mathcal{C}_5}) = -i\int_{\infty}^{\delta} dz \, \frac{e^{-|k|z}}{(z^2-\delta^2)^{\alpha/2}e^{i \pi \alpha /2} + \ell^{\alpha}_{\nu}} \nonumber \\ &-i\int^{\infty}_{\delta} dz \, \frac{e^{-|k|z}}{(z^2-\delta^2)^{\alpha/2}e^{-i\pi \alpha /2} + \ell^{\alpha}_{\nu}} \nonumber \\ &= 2 \int_{\delta}^{\infty} dz \, e^{-|k|z} \nonumber \\ & \times  \frac{\sin(\pi \alpha /2)(z^2-\delta^2)^{\alpha/2}}{(z^2-\delta^2)^{\alpha} + 2(z^2-\delta^2)^{\alpha/2} \,\ell^{\alpha}_{\nu}\cos(\pi \alpha /2) + \ell^{2\alpha}_{\nu}},
\end{align}
where we have changed variables to $z = -i r$ and used the fact that $\mathcal{I}_{\mathcal{C}_5} = \mathcal{I}_{\mathcal{C}_3}^*$. Now we can safely take the limit $\delta \rightarrow 0$. Using Cauchy's residue theorem and also redefining $z \rightarrow \ell_{\nu} z$, we have
\begin{align} \label{g_k_appendix}
    \hat{g}(k)  &= \frac{\varepsilon}{2 \pi \kappa L} \lim_{\delta \to 0} \, \mathcal{I}_{\mathcal{C}_1} = \frac{\varepsilon}{2 \pi \kappa L} \lim_{\delta \to 0} \Big[ \, -(\mathcal{I}_{\mathcal{C}_3} + \mathcal{I}_{\mathcal{C}_5}) \Big] \nonumber \\ &= \frac{\varepsilon \ell_{\nu}}{2 \pi \nu L} \int^{\infty}_{0} dz \, \frac{\sin(\pi \alpha /2) \, z^{\alpha}}{z^{2\alpha} + 2z^{\alpha}\cos(\pi \alpha /2)+1} e^{-|k| \ell_{\nu}z},
\end{align}
which is \eqref{gk_branch}. For $k > 0$, $\mathcal{C}_1$ can be closed by making a contour in the lower half plane, yielding the same expression \eqref{g_k_appendix}. Integrating this expression in $k$ and multiplying by $2 \nu L$ immediately yields \eqref{dissp_k_alpha}.

\subsection{Derivation of \texorpdfstring{\eqref{flux_k_alpha}}{(67)} }

For the $k$ flux, combining \eqref{g_r} and \eqref{kflux_g}, we have
\begin{equation} \label{flux_appendix}
    L \int ds \, \hat{\Gamma}^k = \frac{i \varepsilon}{2 \pi} \int dr \frac{|r|^{\alpha}}{r} \frac{e^{-ikr}}{|r|^{\alpha} + \ell^{\alpha}_{\nu}}.
\end{equation}
This integral can be equated to one where the exponential term is decaying rather than oscillatory by again substituting $|r|^{\alpha} \rightarrow (r^2+\delta^2)^{\alpha/2}$ and then taking $\delta \rightarrow 0$. The method from Appendix \ref{det_calc_subsec1} can be applied again, resulting in \eqref{flux_k_alpha}. The main difference is that this time, the auxillary expansion in $\delta$ introduces a pole at $r = 0$ in the integrand of \eqref{flux_appendix}, making it necessary to deform the contour along the real line. However, this residue contribution vanishes when $\delta \rightarrow 0$, so whether the contour is deformed below or above the pole does not change the final answer.

Given the expression \eqref{flux_k_alpha}, we can take its asymptotics for $k \ell_{\nu} \ll 1$, as analyzed in Section \ref{const_flux}. In this limit, the inertial-range flux is approximately equal to the rate of $\delta C_2$ injection. Changing variables to $u = z^{\alpha}$, we have
\begin{align} \label{flux_k_alpha_IR}
   &  L \int ds \, \hat{\Gamma}^k  \nonumber  \\ & \simeq \mathrm{sgn}(k) \, \frac{\varepsilon \, \sin(\pi \alpha /2) }{\pi \alpha} \int^{\infty}_{0}  \, \frac{du}{u^2 + 2u \, \cos(\pi \alpha /2)+1} \nonumber \\ & = \mathrm{sgn}(k) \, \frac{\varepsilon \, \sin(\pi \alpha /2) }{\pi \alpha} \frac{1}{\sin(\pi \alpha /2)} \frac{\pi \alpha}{2} \nonumber \\  & = \mathrm{sgn}(k) \, \frac{\varepsilon}{2},
\end{align}
where the $u$ integral was done using \cite{gradshteyn2014table}, formula 3.252(1). Note this result also implies that \eqref{dissp_k_alpha} approximately satisfies $\mathcal{\hat{D}}(k) \simeq \varepsilon$ when $k \ell_{\nu} \gg 1$.

\subsection{Expressions for \texorpdfstring{$\alpha = 1$}{alpha = 1} case}

Finally, we note that when $\alpha = 1$, $\hat{g}(k)$, $\mathcal{\hat{D}}(k)$, and $L \int ds \, \hat{\Gamma}^k$ can all be expressed in terms of known special functions. For the sake of completeness, we give these expressions here. 

For $\hat{g}(k)$, using \cite{abramowitz1964handbook}, formulas 5.2.7 and 5.2.13, we have
\begin{align}
    & \hat{g}(k) = \frac{\varepsilon \ell_{\nu}}{2 \pi \nu L} \int^{\infty}_{0} dz \, \frac{z \, e^{-|k| \ell_{\nu}z}}{z^2+1} \nonumber \\ &= -\frac{\varepsilon \ell_{\nu}}{2 \pi \nu L}\left[\cos\big(|k| \ell_{\nu}\big) \textrm{Ci}\big(|k| \ell_{\nu}\big) + \sin\big(|k| \ell_{\nu}\big) \textrm{si}\big(|k| \ell_{\nu}\big) \right],
\end{align}
where $\textrm{Ci}(z)$ and $\textrm{si}(z)$ are cosine and sine integral functions \cite{abramowitz1964handbook}, respectively.

For $\mathcal{\hat{D}}(k)$, using \cite{abramowitz1964handbook}, formulas 5.2.6 and 5.2.12, we have
\begin{align} \label{dissp_k_alpha1}
    &\mathcal{\hat{D}}(k) = \frac{2 \varepsilon }{ \pi} \int^{\infty}_{0}  dz \, \frac{1-e^{-k \ell_{\nu}z}}{z^{2} + 1} \nonumber \\ &= \varepsilon \left\{1 - \frac{2}{\pi}\left[\sin\left(k \ell_{\nu}\right) \textrm{Ci}\left(k \ell_{\nu}\right) - \cos\left(k \ell_{\nu}\right) \textrm{si}\left(k \ell_{\nu}\right)\right] \right\}.
\end{align}
When $k\ell_{\nu} \ll 1$, \eqref{dissp_k_alpha1} has the series
\begin{equation}
    \mathcal{\hat{D}}(k) = \varepsilon\left\{\frac{2}{\pi}\left[1 - \gamma - \log (k\ell_{\nu})\right]k\ell_{\nu} + \mathcal{O}\left((k\ell_{\nu})^2\right) \right\},
\end{equation}
which vanishes as $k\ell_{\nu} \rightarrow 0$. Here, $\gamma$ is the Euler–Mascheroni constant. Note that the finite-$k\ell_{\nu}$ corrections are linear and logarithmic, in contrast to the $\alpha = 2$ case \eqref{dissp_k_alpha2}, where the first-order correction (after Taylor expanding the exponential) is just linear in $k\ell_{\nu}$. When $k\ell_{\nu} \gg 1$, \eqref{dissp_k_alpha1} has the series
\begin{equation}
    \mathcal{\hat{D}}(k) = \varepsilon \left[1-\frac{2}{\pi} \frac{1}{k\ell_{\nu}} + \mathcal{O}\left((k\ell_{\nu})^{-3}\right) \right]. 
\end{equation}

Likewise, the integrated $k$ flux \eqref{flux_k_alpha} is
\begin{align}
& L \int ds \, \hat{\Gamma}^k = \mathrm{sgn}(k) \, \frac{\varepsilon }{ \pi} \int^{\infty}_{0} dz
\, \frac{e^{-|k| \ell_{\nu} z}}{z^{2} + 1} \nonumber \\ & = \mathrm{sgn}(k) \, \frac{\varepsilon }{
\pi} \left[\sin\left(|k| \ell_{\nu}\right) \textrm{Ci}\left(|k| \ell_{\nu}\right) - \cos\left(|k|
\ell_{\nu}\right) \textrm{si}\left(|k| \ell_{\nu}\right)\right].
\end{align}
When $k \ell_{\nu} \ll 1$, this expression has the series
\begin{align} \label{series_k_flux_alpha_1}
L \int ds \, \hat{\Gamma}^k = \mathrm{sgn}(k) \, \frac{\varepsilon }{2} \bigg\{1 -
\frac{2}{\pi}\big[ & 1 - \gamma - \log (|k|\ell_{\nu})\big]|k|\ell_{\nu} \nonumber \\ & +
\mathcal{O}\left((|k|\ell_{\nu})^2\right) \bigg\},
\end{align}
in agreement with \eqref{flux_k_alpha_IR} as $k \ell_{\nu} \rightarrow 0$.

\section{Closed-form expressions of inertial-range spectrum and fluxes for \texorpdfstring{$\alpha = 1$}{alpha = 1}} \label{alpha1_calc_section}

In Sections \ref{PD_section} and \ref{Flux_analysis}, we plotted the inertial-range spectrum and its corresponding vector flux, respectively, for $\alpha = 1$. For this value of $\alpha$, the spectrum and fluxes have simple closed forms, which we derive in this appendix.

When $\alpha = 1$, \eqref{phi} reduces to
\begin{equation} \label{phi_self_similar_alpha_1}
\phi(y) = \tilde{\sigma} \, e^{-(1 - i)y/\sqrt{3}},
\end{equation}
where we have absorbed all order-unity constants into $\tilde{\sigma}$, where $\tilde{\sigma} = 3^{1/4} 2^{-3/4} \sqrt{\pi} \, \sigma$. In \eqref{Fks}, therefore, we have
\begin{align}
     & 2 \operatorname{Re} \int_{0}^{\infty} dy \, e^{-i\xi y} \phi(y)  \nonumber \\ &= 2 \tilde{\sigma} \operatorname{Re} \int_{0}^{\infty} dy \, e^{-y/\sqrt{3}}e^{-iy\left(\xi-1/\sqrt{3}\right)}
    \nonumber \\ &= 2 \sqrt{3} \, \tilde{\sigma}  \, \frac{1}{3\xi^2-2\sqrt{3} \xi + 2}.
\end{align}
The constant $\tilde{\sigma}$ can be found via the flux constraint \eqref{flux_constraint}:
\begin{align}
& \frac{1}{2} = 2 \sqrt{3} \, \tilde{\sigma}  \, \int^{+\infty}_{-\infty}  d \xi \, \frac{\xi}{3\xi^2-2\sqrt{3} \xi + 2} =  2 \sqrt{3} \, \tilde{\sigma} \, \frac{\pi}{3} \nonumber \\ & \implies \tilde{\sigma} = \frac{\sqrt{3}}{4 \pi}.
\end{align}
Therefore, using \eqref{xi} and \eqref{Fks}, the spectrum is
\begin{align} \label{spec_alpha_1}
    \hat{F}(k,s) &= \frac{3 \varepsilon}{2 \pi L \kappa} \, s^{-3} \frac{1}{3\xi^2-2\sqrt{3} \xi + 2} \nonumber \\ &= \frac{3 \varepsilon}{2 \pi L} \, \frac{1}{3k^2 -2 \sqrt{3 \kappa} \, k s^{3/2} + 2\kappa s^3 }.
\end{align}
This expression clearly satisfies the asymptotics given by~\eqref{asymptotics_spec_twiddle}, and the results for the 1-D spectra \eqref{spec_1D_k_twiddle} and \eqref{spec_1D_s_twiddle} apply.

The $s$ flux \eqref{gamma_s} is then simply
\begin{equation} \label{s_flux_a1}
    \hat{\Gamma}^s = k \hat{F} = \frac{3 \varepsilon}{2 \pi L} \, \frac{k}{3k^2 -2 \sqrt{3 \kappa} k s^{3/2} + 2\kappa s^3 }.
\end{equation}
The $k$ flux \eqref{gamma_k}, using \eqref{gamma_k_phi}, is
\begin{align} \label{k_flux_a1}
\hat{\Gamma}^k & = -2 \, \varepsilon \, L^{-1} s^{-1} \operatorname{Im} \int^{\infty}_{0} dy \, e^{-i \xi y} \phi(y). \nonumber \\ & = -\frac{\sqrt{3} \, \varepsilon}{2 \pi L s} \operatorname{Im} \int_{0}^{\infty} dy \, e^{-y/\sqrt{3}}e^{-iy\big(\xi-1/\sqrt{3}\big)} \nonumber \\ &= \frac{3 \varepsilon}{ 2 \pi L s} \frac{\sqrt{3} \, \xi-1}{3\xi^2-2\sqrt{3} \xi + 2} \nonumber \\ &= \frac{3 \varepsilon}{ 2 \pi L} \frac{\sqrt{3 \kappa} \, k s^{1/2}-\kappa s^2}{3k^2 -2 \sqrt{3 \kappa} \, k s^{3/2} + 2\kappa s^3 }.
\end{align}

Finally, we show that the components of the shell-averaged flux \eqref{Gamma_shell} for $\alpha = 1$ are positive-definite. For the $s$ flux, using \eqref{s_flux_a1} and \eqref{Gamma_shell}, we have
\begin{align}
& \hat{\Gamma}^s(k,s) + \hat{\Gamma}^s(-k,s) \nonumber \\ &= \frac{3 \varepsilon k}{2 \pi L \kappa s^3} \left(\frac{1}{3\xi^2-2\sqrt{3} \xi + 2}-\frac{1}{3\xi^2+2\sqrt{3} \xi + 2}\right) \nonumber \\ &= \frac{3 \varepsilon k}{2 \pi L \kappa s^3} \frac{4 \sqrt{3} \xi}{9 \xi^4 + 4} \geq 0.
\end{align}
For the $k$ flux, using \eqref{k_flux_a1} and \eqref{Gamma_shell}, we have
\begin{align}
& \hat{\Gamma}^k(k,s) - \hat{\Gamma}^k(-k,s) \nonumber \\ &= \frac{3 \varepsilon}{ 2 \pi L s} \left(\frac{\sqrt{3} \, \xi-1}{3\xi^2-2\sqrt{3} \xi + 2}+\frac{\sqrt{3} \, \xi+1}{3\xi^2+2\sqrt{3} \xi + 2}\right) \nonumber \\ &= \frac{3 \varepsilon}{ 2 \pi L s} \frac{6 \sqrt{3} \xi^3}{9 \xi^4 + 4} \geq 0.
\end{align}

\section{Characterization of 2-D fluxes for \texorpdfstring{$\alpha < 2$}{alpha < 2} cases} \label{phases_k_flux}

\begin{figure}
	\centering
	\includegraphics[width=1\linewidth]{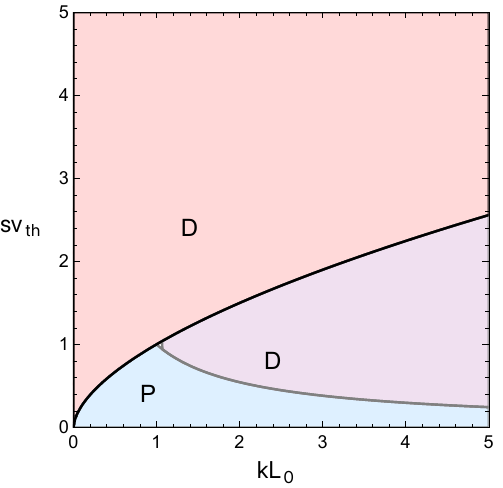}
	\caption{\label{fig:a34_pd} Cartoon diagram depicting the regions where the flux is phase-mixing-dominated (labeled `P' and colored blue) versus diffusion-dominated (labeled `D' and colored red), for $\alpha = 3/4$. The black curve is the critical-balance curve \eqref{CB} for this value of $\alpha$, $s v_{\mathrm{th}} \sim (k L_0)^{7/12}$. The gray curve is \eqref{flux_line} for these parameters, $s v_{\mathrm{th}} \sim (k L_0)^{-7/8}$. We have colored the region bounded above by the critical-balance curve and below by \eqref{flux_line} as purple to emphasize that it is still a region where phase mixing is locally dominant, viz., \eqref{ratio_time_scale} is large, despite the flux being diffusion-dominated there.}
\end{figure}

In this section, we characterize the (non-local) 2-D fluxes for the $\alpha < 2$ cases. The asymptotics of the (normalized) ratio of fluxes \eqref{flux_alpha_ratio} are
\begin{equation} \label{asymptotics_flux_ratio}
 R \sim
\begin{cases} 
      -\left(k L_0\right)^{-1} \, (s v_{\mathrm{th}})^{(5 - \alpha)/(\alpha + 1)}, \qquad \qquad & \xi \ll 1, \\
      (k L_0)^{1-\alpha} \, (s v_{\mathrm{th}})^{(2 \alpha - 1)/(\alpha + 1)}, \qquad \qquad & \xi \gg 1.
   \end{cases}
\end{equation}
Unlike in the Batchelor regime, the ratios in \eqref{asymptotics_flux_ratio} are no longer functions solely of $\xi$, and,  therefore, the regions where $R$ is small or large are not necessarily divided by the critical-balance curve \eqref{CB}. Note that we focus our analysis on the top right quadrant of the $(k,s)$ plane, but our arguments can be extended to the whole plane.

For $\xi \ll 1$, above the curve \eqref{CB}, which is the diffusion-dominated region for $\alpha = 2$, the $s$ and $k$ flux are comparable along the curve
\begin{equation} \label{xi_small_flux_line}
    s v_{\mathrm{th}} \sim (k L_0)^{(\alpha + 1)/(5-\alpha)}.
\end{equation}
At sufficiently large $k L_0$, this curve falls below the critical-balance curve \eqref{CB}, outside the $\xi \ll 1$ regime, and hence $\hat{\Gamma}^{k}$ is dominant over $\hat{\Gamma}^{s}$ when $\xi \ll 1$ for all $\alpha$.

For $\xi \gg 1$, below the curve \eqref{CB}, which is the phase-mixing-dominated region for $\alpha = 2$, the $s$ and $k$ flux are comparable along the curve
\begin{equation} \label{flux_line}
   s v_{\mathrm{th}} \sim (k L_0)^{(\alpha^2-1)/(2 \alpha - 1)}. 
\end{equation}
There are now various regimes with different behaviors.

When $1 < \alpha < 2$, the flux is phase-mixing-dominated (diffusion-dominated) below (above) the curve \eqref{flux_line}. At sufficiently large $k L_0$, the curve \eqref{flux_line} falls below the critical-balance curve \eqref{CB}, effectively widening the diffusion-dominated region from the $\xi \ll 1$ region down to the curve \eqref{flux_line}. Note that for this range of $\alpha$, \eqref{flux_line} is concave, same as \eqref{CB}.

When $\alpha = 1$, \eqref{flux_line} is horizontal in $k L_0$, and so the flux is only phase-mixing-dominated for $s v_{\mathrm{th}} \ll 1$, irrespective of $k$. We plotted this case in Fig.~\ref{fig:a1_flux_ks_alpha2_alpha1}(b).

When $1/2 < \alpha < 1$, $s v_{\mathrm{th}}$ in \eqref{flux_line} is a decreasing function of $k L_0$, shrinking the phase-mixing-dominated region further. As an example, we plot a diagram for the case $\alpha = 3/4$ in Fig.~\ref{fig:a34_pd}. Note that when $\alpha < -1 + \sqrt{3} \simeq 0.732$, the dependence of $s v_{\mathrm{th}}$ on $k L_0$ in~\eqref{flux_line} is steeper than $(k L_0)^{-1}$, so the area of the phase-mixing-dominated region becomes finite.

When $\alpha = 1/2$, the flux is phase-mixing-dominated only when $k L_0 \ll 1$, irrespective of $s$. Note that $k$ must satisfy $k L_E \gg 1$ to be in the inertial range (and $\xi \gg 1$ for these asymptotics to hold), so the phase-mixing-dominated region is likely nonexistent.

When $0 < \alpha < 1/2$, the curve \eqref{flux_line} is convex. The phase-mixing-dominated region is bounded below by \eqref{flux_line} and above by \eqref{CB}, extending from $k = 0$ to the intersection of \eqref{flux_line} and \eqref{CB}, viz., at $k L_0 = 1$. The region where the flux is phase-mixing-dominated is not asymptotic, i.e., everywhere in the $\xi \gg 1$ region, in the limit $\xi \rightarrow \infty$, either in the limit $k L_0 \rightarrow \infty$ and/or $s v_{\mathrm{th}} \rightarrow 0$, the flux is diffusion-dominated. Therefore, in this case, the only part of the $(k,s)$ plane where the flux is phase-mixing dominated is along the curve $\xi = \xi_{\alpha}$, where the $k$ flux vanishes.

\bibliography{paper}

\end{document}